\documentclass{ieeeaccess}

\usepackage{cite}
\usepackage{url} % For proper URL formatting
\usepackage{graphicx} % For including images
\usepackage{booktabs} % For better table formatting
\usepackage{caption} % For better caption control
\usepackage{float} % For better float positioning
\usepackage{amsmath}   % 如果你用过 \text{} 或对齐环境，也需要
\usepackage{amssymb}   % 提供 \mathbb
\usepackage{geometry}
\usepackage{subcaption} % 导言区添加
\usepackage{caption}
\DeclareCaptionLabelFormat{appendixfigure}{Appendix Figure \arabic{figure}}
\begin{document}
	\doi{}
	\dataDaDefesa{20{\slash}09{\slash}2025}
	\title{Scaling Law in LLM Simulated Personality: More Detailed and Realistic Persona Profile Is All You Need}
	
	\author{\uppercase{Yuqi Bai}\authorrefmark{1},
		\uppercase{Tianyu Huang}\authorrefmark{2},
		\uppercase{Kun Sun}\authorrefmark{1},
		\uppercase{Yuting Chen}\authorrefmark{2}}
	
	\address[1]{Department of Computer and Information Engineering, Hebei Petroleum University of Technology, Chengde, China (e-mail: yuqi.bai8888@gmail.com; heishuojinjsj@qq.com)}
	\address[2]{School of Computer Science \& Technology, Beijing Institute of Technology, Beijing, China (e-mail: huangtianyu@bit.edu.cn; chenyuting@bit.edu.cn)}
	
	\begin{abstract}
		This research focuses on using large language models (LLMs) to simulate social experiments, exploring their ability to emulate human personality in virtual persona role-playing. The research develops an end-to-end evaluation framework, including individual-level analysis of stability and identifiability, as well as population-level analysis called progressive personality curves to examine the veracity and consistency of LLMs in simulating human personality. Methodologically, this research proposes important modifications to traditional psychometric approaches (CFA and construct validity) which are unable to capture improvement trends in LLMs at their current low-level simulation, potentially leading to premature rejection or methodological misalignment. The main contributions of this research are: proposing a systematic framework for LLM virtual personality evaluation; empirically demonstrating the critical role of persona detail in personality simulation quality; and identifying marginal utility effects of persona profiles, especially a Scaling Law in LLM personality simulation, offering operational evaluation metrics and a theoretical foundation for applying large language models in social science experiments
	\end{abstract}
	
	
	\titlepgskip=-15pt
	
	\maketitle

	\section{Introduction}
	Our research centers on employing large language models (LLMs) for the simulation of social experiments\cite{ref24,ref25,ref26}. In this setting, LLMs engage in role-playing to construct virtual human individuals and populations, which in turn serve as proxies for human society in experimental investigations. Because personality is a foundational dimension of human attributes, the capacity of LLMs to simulate personality becomes a cornerstone for modeling human individuals and collectives. Under this framework, research on LLM personality does not concern the intrinsic personality of the models themselves \cite{ref1,ref2,ref3}. Instead, it examines how LLMs, when guided by persona profiles, can simulate and approximate the personalities of human individuals and populations through role enactment. Throughout this paper, the notions of virtual, simulated, and simulation are employed exclusively in reference to this role-playing context.
	
	A key hypothesis about the emergent abilities of LLMs holds that, by training on massive textual corpora, these models have internalized a latent ``world model'' within their parameters—one that encodes probabilistic distributions of entities, events, and human behaviors in the form of conditional probability structures \cite{ref4,ref5,ref6}. While this hypothesis is true only to a certain extent, its primary value lies in providing a constructive interpretive framework. Within this framework, all LLM inferences and generations can be understood as processes supported by these probabilistic structures. Accordingly, LLMs are not only expected to generate linguistically coherent text, but also to infer complex social contexts and individual characteristics from limited cues, thereby opening new possibilities for conducting virtual experiments in the social sciences.
	
	We adopt the notion of a world model as a unifying methodological framework for our analysis. The first component of our framework is a synthesis method that combines census-based population sampling with LLM-driven conditional generation, enabling the construction of virtual persona profiles. These profiles are grounded in real statistical distributions while enriched with detailed social attributes and measurable psychological characteristics. Within this process, demographic data provide the factual and conditional foundation, whereas the LLM's generation of fine-grained persona descriptions represents the inference process guided by the probability structures of the world model.
	
	The second component involves presenting an LLM with a persona profile and prompting it to simulate personality. In this step, the persona profile serves as the factual basis from which personality traits are inferred. Anchored in this two-stage narrative structure, we propose an end-to-end framework for assessing the capacity of LLMs to conduct social experiment simulations. The evaluation spans from individual-level statistical analyses to population-level comparisons of personality distributions.
	
	At the population level, we specifically examine whether the Big Five personality trait curves across age in virtual populations align with empirical human survey data. Given that previous studies have found that the distribution curves of personality traits along the age dimension show consistency across different countries, this method provides a rigorous benchmark for validating the authenticity of LLM-based social simulations.
	
	\subsection{Methodological Issues: CFA and Measurement Invariance}
	On CFA model fit and measurement invariance in psychological assessment
	Researches of LLM personality predominantly rely on human personality measurement instruments, and this reliance is self-evidently justified. As discussed above, the context of LLM personality research lies in social simulation experiments where LLMs mimic human behavior. Methodologically, this very alignment necessitates the use of measurement approaches consistent with human psychological assessment, which remains the standard practice in this research domain  \cite{ref7}.
	
	At the same time, numerous studies apply CFA validation. Specifically, CFA is conducted on datasets of virtual persona personality traits generated by LLMs, with model fit indices—particularly measurement invariance—used as a purported indicator of the LLM's capacity to shape virtual personalities, and in some cases, as evidence of whether an LLM possesses or might possess a human-like psychological structure.
	
	However, a substantial body of research, particularly cross-population studies, demonstrates that even for human data, achieving adequate CFA fit and measurement invariance is challenging. Fit and invariance metrics are extremely sensitive to differences across human populations, including across racial and cultural groups.
	
	Regardless of the an LLM’s capability to approximate real human personality—or even if it already generates highly realistic approximations—it is difficult to conceive that such models could attain a level of fidelity exceeding the variability observed among human populations themselves. Only under such extreme fidelity would CFA validation potentially succeed. To date, no method has reached this level of realism. Identifying a pathway toward such fidelity constitutes a core mission of the field and remains a central challenge. Pursuing this objective requires an evaluation framework capable of guiding the search for this pathway; In other words, such a framework should not function like a microscope that scrutinizes whether LLM-simulated personalities have already reached an extreme proximity to human personalities; rather, it should operate like a telescope, observing the direction and trajectory of change under various influencing factors. Only with a sufficiently broad field of vision can such an evaluation be achieved; clearly, CFA is not such a framework. Applying CFA to LLM-generated personality measurement data, and interpreting convergence, factor loadings, or measurement invariance as indicators of model capability, is both trivial and irrelevant.
	
	Some research in this domain has employed measures of construct validity rather than CFA. Although these approaches are somewhat less stringent than CFA, we contend that, as measures of human personality, they are still methodologically misaligned and misapplied. Consequently, conclusions drawn from such methods warrant careful reconsideration.
	
	It is important to highlight the core issue in this field: identifying methods and pathways that enable continuous improvement in LLMs' capacity to simulate personality. Within this context, an appropriate evaluation method is required—one capable of detecting trends in the development of LLM personality simulation, even when performance remains at a low level. Highly sensitive "microscopic" instruments such as CFA, or less sensitive construct validity metrics, generally lack the necessary scope to observe such developmental trajectories. They either dismiss the potential of LLMs prematurely or steer research toward methodologically irrelevant conclusions.
	
	What is therefore needed is an evaluation framework that simultaneously accounts for sensitivity and developmental progress: it should capture both the current shortcomings of LLMs at low performance levels and the incremental improvements over time. In other words, the goal of evaluation should not be to determine whether human-level performance has already been attained, but to record and characterize the trajectory toward that level.
	
	In our research, we employ measurement scales while explicitly avoiding both CFA and construct validity methods \cite{ref7,ref8,ref9,ref10,ref11,ref12,ref13,ref14}.
	
	\subsubsection{Additional Notes on Personality Scales and Test Evaluation}
	While the use of CFA or construct validity metrics inevitably introduces certain challenges, inappropriate attempts to address these challenges may create additional complications. For instance, some approaches explore whether LLMs lack "real personality" and, based on this premise, reject the use of personality scales altogether, or attempt to devise non-human personality measurement methods.
	
	Debates over whether LLMs possess "real personality" effectively undermine the foundational premise of the Turing test, diverge from empirical scientific methodology, and risk falling into metaphysical speculation. Even granting metaphysical perspectives, once we recognize that LLMs possess no intrinsic mind and are not intended to instantiate a new non-human "personality," but only to simulate human personality, it is for this very reason that the methods of personality assessment should all the more strictly adhere to those applied to humans.
	
	This requirement is directly dictated by the core issue of the field. In social experiments involving human personality simulation, the methods for personality assessment must mirror those used for humans. What requires reconsideration is not the measurement instruments themselves, but rather the methods for evaluating LLM performance on these assessments. In general, applying human psychometric instruments to LLM personality simulation is not intrinsically erroneous. Rather, it is a matter of contextualized judgment under the core objectives of the field. Personality scales are both correct and essential, a necessity arising directly from the broader context of personality simulation. Similarly, CFA and construct validity methods are not incorrect, but they lack the capacity to capture the developmental trajectory of LLMs as they approximate human personality.
	
	\begin{figure*}[htbp]
		\centering
		\includegraphics[width=0.8\textwidth]{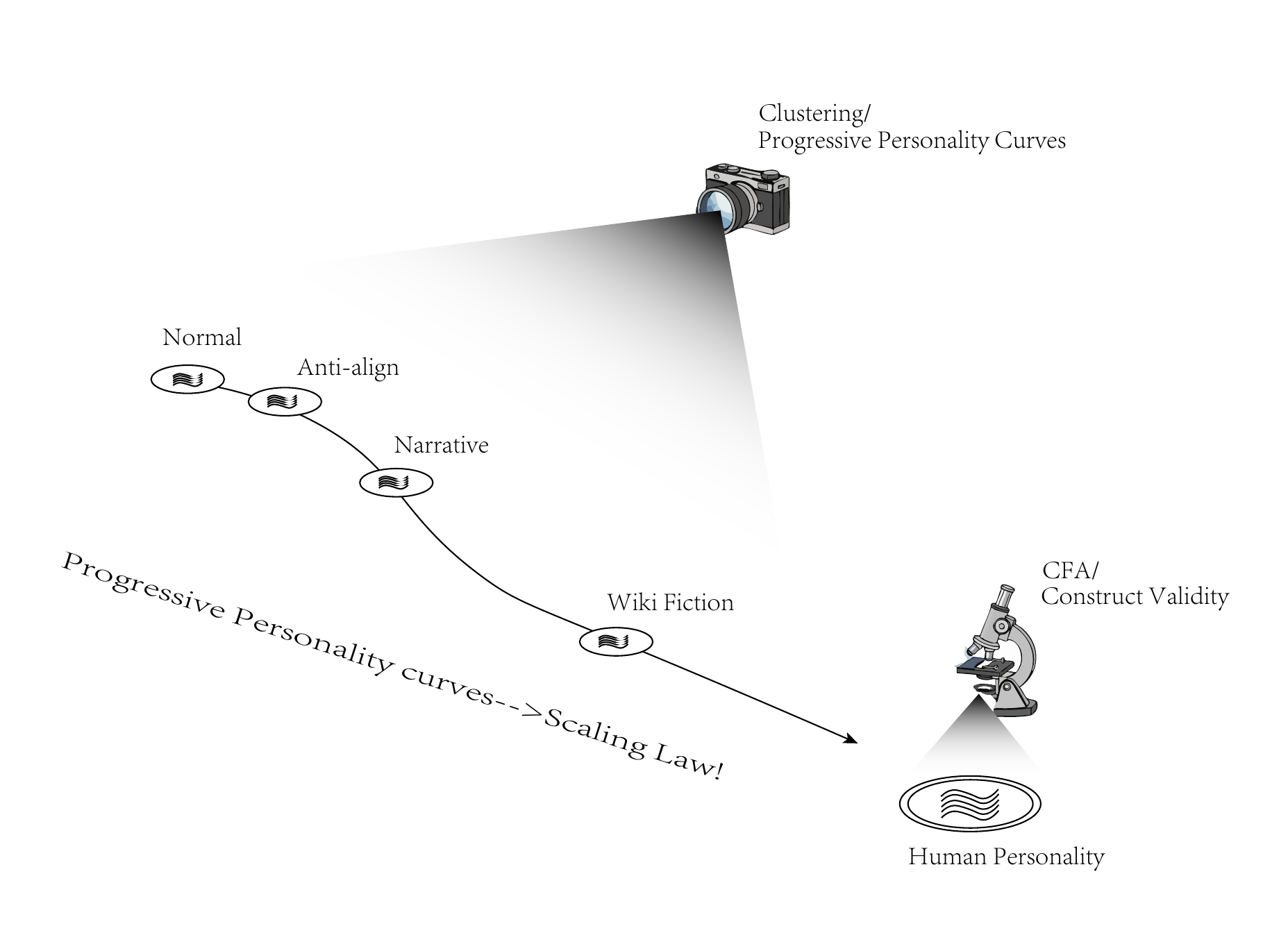}
		\caption{Methodological Scope Comparison: Traditional approaches (CFA and construct validity) focus narrowly on microscopic human personality space, while our engineering analytics approach captures the full developmental trajectory from current LLM capabilities toward human-like personality simulation.}
		\label{fig:methodology_comparison}
	\end{figure*}

	\subsection{Research Content}
	1) Research Framework
	Building on the world model narrative described above, we establish a comprehensive research framework, structured as follows:
	1.\ Virtual persona synthesis
	2.\ Virtual persona personality testing
	3.\ Result evaluation
	Within this framework, we conduct two types of experimental procedures:
	1.\ Using population demographic data as input for virtual persona synthesis → generating virtual personas → personality testing → conducting evaluation
	2.\ Using alternative persona data sources (wikidata fictional characters) to drive virtual persona generation → personality testing → conducting evaluation
	
	The first type of experiment seeks to investigate the two-step generative and inferential capabilities of the LLM world model, while the second type focuses on the second-step generation and inference under conditions where persona profiles with a certain degree of realism and detail are already available.
	
	The scenario corresponding to the first type of experiment is evidently more ideal: it fully engages the LLM world model's generative and inferential abilities, providing only the statistical features of population composition, while the LLM synthesizes persona profiles to drive personality simulation. In most social science experiments, sample selection is a fundamental and critical step, directly influencing experimental validity and scientific rigor. Nevertheless, access to real population data is often limited by resource constraints and privacy considerations. Generating skeletal persona profiles based on statistical models derived from real census data provides a simplified yet demographically representative foundation. Using carefully designed prompts, the LLM conditionally generates detailed personas, enriching each with life experiences, interests, and behavioral characteristics, ultimately producing complete persona profiles that are both statistically realistic and diverse. If the LLM is sufficiently powerful in this first-step generative inference, this approach not only mitigates privacy risks associated with real data but also effectively addresses resource limitations.
	
	The second type of experiment pertains to application scenarios that draw on authentic and fine-grained persona data sources, a setting of substantial value in its own right. From a research standpoint, this design enables an independent assessment of the LLM's capacity to generate and infer personality traits under conditions where persona profiles are already specified.
	
	To illustrate this mechanism from a Bayesian viewpoint, we employ the metaphor of an "experienced detective": given basic demographic information about a suspect (e.g., age, gender, occupation), the detective, drawing upon years of accumulated social experience and a deep understanding of human behavior, can plausibly infer the suspect's life details, psychological traits, and behavioral patterns. Analogously, an LLM, functioning as an agent for language comprehension and generation and guided by its internalized world model, can infer and generate virtual persona profiles that are both richly detailed and internally consistent from the provided conditional information. In this analogy, generating detailed persona profiles from demographic skeleton data corresponds to the first step of detective-like reasoning, while inferring personality traits from these profiles constitutes the second step. Assessing the LLM's performance along this reasoning chain effectively probes the fidelity and applicability of its internalized world model within the domain of human behavior.
	
	2) Individual-Level Evaluation
	Personality traits constitute the fundamental dimensions of human behavior. In this research, the Big Five, a set of widely recognized psychological measures with established reliability and validity, are adopted as the primary indicators for assessing the plausibility of virtual personality. Each generated virtual persona is employed for LLM role-playing, through which it participates in the Big Five personality assessment.
	
	At the individual level, statistical techniques such as Mahalanobis distance and cluster analysis are applied to rigorously assess the personality characteristics of virtual personas, thereby evaluating their plausibility, stability, and internal consistency.
	
	3) Population-Level Evaluation
	At the population level, evaluation focuses on the plausibility and stability of the overall personality distribution within the virtual population. This is achieved by comparing the age-related trajectories of the Big Five personality traits in the virtual population with the corresponding trajectories observed in empirical ground-truth human survey data \cite{ref22}. In psychology, age-related changes in personality traits are recognized as key features that are stable across population groups \cite{ref15,ref16,ref17,ref18,ref20}. Accordingly, this comparison serves as an end-to-end benchmark for assessing the LLM's ability to simulate human populations accurately.
	
	At the population level, beyond the LLM-generated persona profiles, we incorporate profiles derived from external data sources to investigate the influence of varying origins and quality of persona data on simulated personality outcomes.
	
	\subsection{Contributions}
	The main contributions of this research are reflected in the following two aspects:
	
	First, we propose a framework for evaluating the personality simulation capabilities of large language models,  with the aim of capturing the trajectory by which LLMs approach human-like personality representations. This framework encompasses:
	\begin{itemize}
		\item Virtual Persona Generation Combining Population Demographics and LLMs: This method constructs skeletal persona profiles that conform to real-world population distributions and employs carefully designed prompts to guide the LLM in generating richly detailed virtual personas with statistical realism. This approach addresses the challenges of limited data resources and privacy protection commonly encountered in traditional social science experiments.
		\item Individual-Level Evaluation Framework: We design and implement an evaluation framework at the individual level, combining Big Five personality assessments with statistical techniques such as Mahalanobis distance and cluster analysis to verify the convergence and stability of personality traits in generated virtual personas. Notably, improvements in convergence and stability and identifiability are observed as the level of detail in persona profiles increases.\textbf{ An important insight revealed in this part is the marginal utility effect of persona profile detail. }.
		\item \textbf{Progressive Personality Curves:}Population-Level Evaluation Based on Age-Related Personality Distributions: We establish a population-level evaluation benchmark by examining the trajectories of personality traits across age in virtual populations and comparing them with empirical human survey data. This systematic evaluation assesses LLM performance in social simulation experiments and provides actionable metrics and an empirical foundation for future LLM-based social science research.
	\end{itemize}
	Second, a series of experiments implemented within this research framework revealed a Scaling Law in LLM-based personality simulation driven by persona profiles. Specifically, as the persona profiles used as prompts become more realistic, richly detailed, and comprehensive, individual-level personality traits of virtual personas exhibit greater stability, convergence, and distinguishability. At the population level, the simulated personalities increasingly align with the statistical characteristics observed in real human populations. Consequently, limitations previously noted in LLMs' performance for this task are progressively alleviated and eventually diminish.
	
	This Scaling Law in virtual persona personality simulation offers a clear and actionable path for addressing the key challenges faced in this domain.
	
	\section{Related Research}
	Psychometric approaches have been widely adopted in research on LLM-based psychological simulations \cite{ref7}, yet they face a range of challenges \cite{ref8,ref9,ref10,ref11,ref12,ref13,ref14}.
	
	In human cross-population personality research, achieving measurement invariance within CFA models remains notoriously difficult. Studies indicate that personality assessments across populations often encounter issues in CFA model fit. For example, \cite{ref15} reported that CFI values for dimensions such as Extraversion and Openness were generally below 0.90 across different countries, while the Neuroticism and Conscientiousness dimensions exhibited relatively better fit. Analysis of IPIP-NEO-120 data from 12 East Asian and Northern European countries showed that full measurement invariance was difficult to attain; nevertheless, partial adjustments enabled meaningful cross-cultural comparisons \cite{ref16}. Similarly, the BFPTSQ scale demonstrated partial invariance at both language and country levels, while the overall structure remained supported \cite{ref3,ref17}. \cite{ref18} further noted that the NEO-FFI achieved only configural invariance across cultural groups, without supporting threshold invariance.
	
	These results provide valuable guidance for evaluating LLM-simulated personalities . Since LLM-simulated personalities remain a considerable distance from real human traits, the suitability of highly sensitive methods, such as CFA and construct validity assessments, for evaluating them is open to question.
	
	In recent years, researchers have explored the use of traditional human personality scales to evaluate large language models (LLMs). Empirical findings, however, reveal significant issues regarding construct validity. \cite{ref9} observed that conventional self-report personality tests exhibit high instability when applied to LLMs: personality scores are extremely sensitive to prompts and item order, and the same model may display markedly different personality profiles under varying conditions, highlighting the inadequacy of human scales for measuring LLM personality.
	
	\cite{ref2} introduced the TRAIT testing tool, employing large-scale multiple-choice assessments to evaluate LLM personality. While results indicate a degree of consistency influenced by training data, prompting methods are limited in eliciting certain traits (e.g., high psychopathy or low conscientiousness), further underscoring the limitations inherent in human-based scales. \cite{ref19}, using latent variable analyses, demonstrated that even when LLMs show high internal consistency and stable factor structures, these outcomes may reflect the effects of training data and prompting strategies rather than genuine psychological characteristics, suggesting that LLM "personality" may constitute a "cognitive ghost" formed during training.
	
	Taken together, these studies suggest that traditional human personality scales lack adequate construct validity for evaluating LLMs and cannot reliably capture their personality traits. Future research should therefore move beyond direct adaptation of human scales and develop novel testing approaches specifically tailored to LLMs, enabling more accurate and model-appropriate personality measurement.
	
	\cite{ref8} identified five key challenges for LLMs in the context of social simulations: diversity, bias, sycophancy, alienness, and generalization. In terms of diversity, LLMs tend to produce outputs that are overly generic or stereotypical, failing to capture the rich variation inherent in human behavior. Concerning bias, models may inherit systematic biases present in their training data, resulting in distorted portrayals when simulating specific population groups. With respect to sycophancy, LLMs can exhibit excessive compliance with experimental protocols or researcher expectations, potentially skewing simulation outcomes toward idealized patterns. Alienness refers to the tendency of models to generate behavior that is inconsistent with human norms, particularly in complex or novel scenarios. Finally, generalization pertains to the models' variable performance across contexts—they may excel in familiar scenarios but behave unpredictably or inconsistently in others. Collectively, these challenges pose significant constraints on the fidelity and reliability of LLM-based human behavior simulations.
	
	\section{Our Method}
	\subsection{Experimental Framework}
	The experimental framework is a general-purpose pipeline that supports the experimental procedures in this research. This research requires a series of experiments to examine and observe various attributes of LLM personality simulation and to explore its influencing factors. The experimental data come from different sources and are organized differently. Sometimes, IPIP-NEO-120 questionnaire simulations are conducted on groups of persona profiles; sometimes, repeated experiments are performed on the same persona profile. Occasionally, synthetic persona profiles are generated and subjected to a full experimental process, while at other times, external persona data sources are introduced indirectly. The experimental framework functions as a customizable personality testing pipeline, supporting the execution of all experiments designed in this research. The resulting data are used to evaluate various attributes of LLM personality simulation and to identify influencing factors. This experimental framework does not include the analysis and evaluation components; it is a pipeline for producing experimental data, consisting of three stages: persona synthesis, IPIP-NEO-120 personality testing, and personality extraction. Based on this foundational framework, data for specific experiments are generated to serve as the foundation for analysis and evaluation. 
	For a more detailed description of the experimental framework, see Appendix 1.
	
	\subsection{Evaluation Method}
	In evaluating the personality test outcomes of virtual personas generated by large language models (LLMs), we deliberately move away from traditional psychometric approaches, instead employing engineering-oriented analytical methods. As outlined in the methodological discussion of the introduction, the primary objective is to capture the trajectory through which LLM-simulated personalities improve and increasingly approximate human-like traits.
	
	\subsubsection{Individual-Level Personality Evaluation}
	Individual-level experiments are critical in the research of virtual persona modeling and personality simulation, as they assess whether LLM-driven characters exhibit psychological consistency and discriminability. In this research, we designed two complementary types of experiments at the individual level: one examining personality stability and convergence, and the other evaluating the identifiability of personalities.
	
	In the personality stability (convergence) experiment, we investigated whether an LLM could maintain stable personality traits over prolonged role-playing scenarios and whether these traits would progressively converge across repeated measurements. To this end, a single persona profile underwent 300 consecutive personality assessments. The Mahalanobis Distance was employed to quantify the distance of each assessment from the distributional center, accounting for correlations among personality dimensions. This approach provides a rigorous, quantitative measure of stability across repeated trials.
	
	The personality identifiability experiment, in contrast, assessed the distinctiveness of different personalities within the personality trait space. For this purpose, two distinct persona profiles were each subjected to 300 personality assessments. The resulting data were combined and subjected to clustering analysis to determine whether distinguishable clusters emerged naturally in trait space. This unsupervised approach effectively captures intrinsic differences between personalities, offering a robust measure of identifiability.
	
	Both experiments incorporated the variable of "persona detail level." We tested not only standard and finely detailed persona profiles but also coarser versions, enabling comparisons of how different levels of detail influence stability and identifiability. This design allows us to explore whether the granularity of persona specifications systematically affects trait compactness and individual discriminability, offering a methodological lens for understanding the role of persona detail in virtual character modeling.
	
	Collectively, these experiments address two central questions: (1) Do LLM-driven virtual characters exhibit stable and convergent personality traits? (2) Can different persona profiles consistently display distinguishable individual characteristics in large-scale personality assessments? The integration of Mahalanobis distance, clustering analysis, and systematic variation in persona detail forms the conceptual and methodological backbone of our experimental design.
	
	\subsubsection{Population-Level Evaluation}
	Building upon the validation of personality stability and identifiability at the individual level, this research extends the analysis to the population level. 
	
	Previous research has demonstrated that the average scores of the Big Five traits across age groups form trajectories that are stable across populations \cite{ref20}, highlighting the cross-population invariance of age-related personality patterns. The cross-population invariance of age-related Big Five trajectories suggests that a common age-personality curve exists across human populations. This curve serves as a reference point for assessing the extent to which LLM-generated population-level personalities resemble real human personality. In our study, we aggregated LLM-simulated personalities, constructed their age-related curves, and compared them against the human benchmark. 
	
	The personality–age curve can be conceptualized as a representative of population-level personality statistics, or even as a proxy for the underlying distribution of personality traits. Accordingly, the extent to which the curve generated from LLM-simulated populations aligns with the empirically observed human curve constitutes a strong indicator of convergence between the simulated and real personality distributions. In the following discussion, we refer to the personality–age curve as the personality curve, and we treat it as synonymous with the population-level distribution of personality traits or their statistical characteristics.
	
	The experimental design addresses three central questions:
	1.\ Consistency: Do LLM-generated populations replicate the age-related trajectories observed in human data?
	2.\ Bias Identification: Should notable deviations from human statistical norms arise, can these be traced to the model's alignment training or inherent generation biases?
	3.\ Authenticity Enhancement: To what extent can the realism and granularity of the virtual population be progressively enhanced—through diverse persona generation strategies, prompt engineering techniques, and the incorporation of external persona data sources (e.g., Wikidata)—so that the simulated personality distributions increasingly mirror the characteristics of actual human populations?
	
	The population-level evaluation comprised four complementary sub-experiments:
	\begin{itemize}
		\item Standard LLM-Generated Personas Experiment: A total of 600 persona profiles were generated by interpolating demographic data and subsequently enriched through LLM-based detail expansion. Personality assessments and personality curve were computed to evaluate the degree to which LLM-generated virtual populations align with real human populations along personality curve.
		\item Bias Mitigation Experiment: To address systematic positive biases identified in the standard generation method, anti-alignment prompts were implemented. These prompts guided the model to produce more realistic and naturalistic outputs, effectively reducing systematic deviations.
		\item Narrative Generation Experiment: A narrative-oriented generation framework, inspired by novel-writing techniques, was employed to encourage LLMs to construct persona profiles in a concrete, story-driven manner. This method aimed to enhance persona complexity and realism, while examining its impact on personality statistical patterns.
		\item Literary Personas Experiment: At the highest level of persona authenticity, literary persona profiles created by human authors (sourced from Wikidata) were directly used as input. This experiment tested whether such high-authenticity personas could further align LLM-generated personality distributions with human population benchmarks.
	\end{itemize}
	
	Notably, in the population-level experiments, we implemented a stepwise progression in persona authenticity and detail, beginning with simplified demographic interpolations, advancing through bias-mitigated and narrative-enriched generations, and culminating in human-authored literary personas. Each stage produced a corresponding personality curve, enabling systematic observation of how simulated population-level distributions evolve under increasingly realistic conditions.
	
	\paragraph{Progressive Personality Curve}
	Building on this framework, we introduce the notion of progressive personality curves, referring to the series of personality curves generated across successive stages of persona sophistication. Rather than treating a personality curve as a static representation of a single simulated population, this concept emphasizes the ensemble of curves and their dynamic trajectory toward empirically observed human benchmarks. Progressive personality curves thus provide both a systematic indicator of convergence and a conceptual tool for quantifying the impact of persona design choices on population-level personality modeling.
	
	\section{Experimental Setup}
	\subsection{Skeletal Persona Sampling Engine Based on Census Data}
	For the construction of the skeletal persona sampling engine, we employed the Adult Income Dataset \cite{ref21}. While this dataset is somewhat dated and its demographic composition deviates considerably from the current population—thus limiting its representativeness of contemporary statistics—these limitations do not compromise the methodological validation pursued in this research. Its widespread adoption in academic research and the extensive availability of Jupyter Notebook resources make it a practically valuable tool for our experiments. The sampling engine supports both random and conditional sampling modes, with conditional sampling being particularly critical for experiments targeting specific subpopulations.
	
	\subsection{Large Language Model}
	To address the primary objective of this research—assessing the capacity of large language models (LLMs) to generate stable and distinguishable personality profiles under varying persona settings—all experimental tasks were executed on a single LLM platform (Deepseek-chat, version 3.0, prior to the 3v.1 upgrade). The focus of the research is not on comparing inherent personality traits across different models; rather, it examines whether, under a controlled and fixed model, prompt-conditioned virtual personas can exhibit individual-level consistency, identifiability, and population-level alignment with statistical patterns observed in real human populations. By treating the model as a constant generative system, this design allows for an evaluation of its ability to simulate diverse, psychologically coherent personality trait expressions in response to varied input conditions. This approach minimizes confounding effects stemming from inter-model differences and provides a more precise assessment of prompt-driven personality construction.
	
	The API was integrated within the Langhian framework, facilitating flexible model invocation and enhancing system adaptability and scalability for future experiments. All requests were made with a default temperature setting of 0.7.
	
	\subsection{Personality Trait Assessment}
	The Big Five Personality Traits framework is a widely accepted theoretical model in personality psychology and has been extensively applied in recent LLM-based personality studies \cite{ref11,ref12,ref13,ref14}. In this research, personality assessment was conducted using the IPIP-NEO-120 questionnaire developed by Professor John A. Johnson. This instrument is a streamlined version of the IPIP-NEO-300 model by Lewis R. Goldberg, and it demonstrates strong reliability, validity, and practical applicability. Implementation involved modifications and local adaptations of existing open-source assessment tools \cite{ref23}.
	
	To mitigate computational costs, we initially attempted to submit all 120 items in a single request to the Deepseek model, but only around 50 responses were returned. Similar limitations were observed with ChatGPT, which could respond to a maximum of 101 items. Consequently, the questionnaire was partitioned into six groups of 20 items each. This approach significantly reduced computational load while ensuring completeness of model responses. The responses from the six groups were subsequently parsed, merged, and processed by the IPIP-NEO assessment engine to yield the final personality trait scores.
	
	\section{Results}
	\subsection{Experimental Framework}
	The synthesis of persona profiles and other relevant results of the experimental framework are provided in Appendix 1.
	
	\subsection{Individual-level Experimental Results}
	In this part of the experiment, we randomly selected five LLM-generated persona profiles and created simplified versions by removing some of their content, which we refer to as the \textit{poor version}. Both the simplified (\textit{poor}) versions and the original, unmodified (\textit{standard}) versions of the persona profiles were then each subjected to 300 repeated personality tests. The resulting data served as the basis for the two individual-level experiments: \textit{stability (convergence)} and \textit{identifiability}. 
	\subsubsection{Stability (Convergence) of Virtual Characters' Personality}
	Objective: To examine whether virtual characters generated via LLM role-playing demonstrate stable and convergent personality traits.
	The incomplete sample counts shown in the figures are due to occasional failures in LLM invocation,, as well as outlier filtering in the statistical analysis. The number of missing samples is generally within the single-digit range and therefore does not have a substantive impact on the statistical results. 
	
	Methodology:
	\begin{itemize}
		\item Baseline Procedure: Each character profile was subjected to 300 successive personality assessments to determine the stability and convergence of personality traits within the LLM-driven role-playing context.
		\item Detail-Level Comparison: Persona profiles with varying levels of descriptive detail were subjected to identical personality assessments to evaluate differences in trait compactness and stability.
	\end{itemize}
	
	Evaluation Metrics:
	\begin{itemize}
		\item Mahalanobis Distance: Quantifies the deviation of each response from the "ideal personality prototype," with lower values indicating closer alignment to the expected personality.
		\item Probability Density Estimation (KDE): Illustrates the central tendency and concentration of responses.
		\item Coefficient of Variation (CV): Captures data dispersion, where smaller values denote greater stability.
		\item Kurtosis: Negative values reflect flatter, light-tailed distributions indicative of stronger normality; values near zero indicate a more uniform distribution, whereas pronounced negative values suggest wider distribution tails.
	\end{itemize}
	
	\begin{figure*}[htbp]
		\centering
		\includegraphics[width=0.8\textwidth]{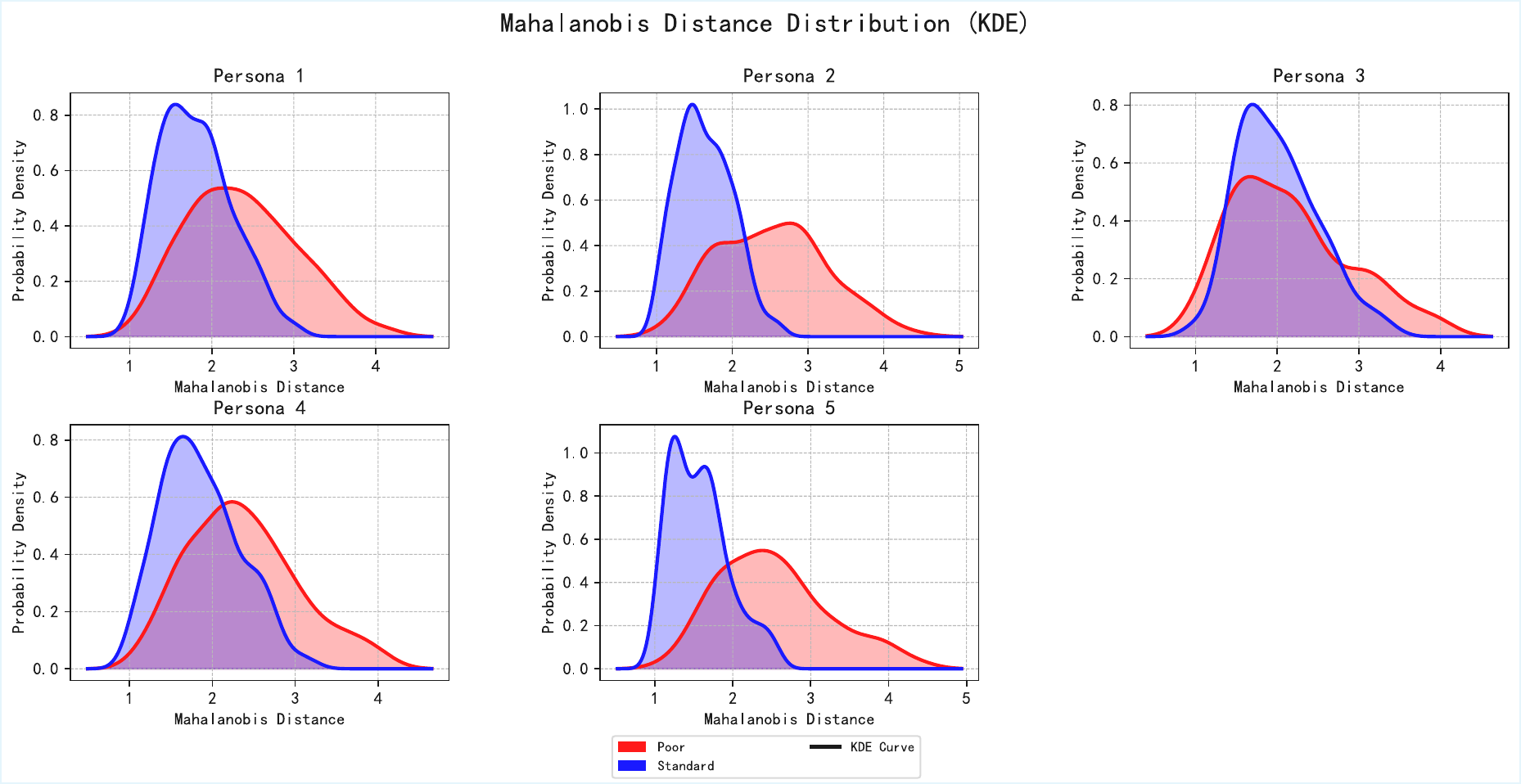}
		\caption{Convergence and stability of LLM simulated personality}
		\label{fig:convergence}
	\end{figure*}
	
	\begin{table*}[!htbp]
		\centering
		\caption{Stability and convergence metrics for LM-simulated personality (VC: Variation Coefficient; SC: Sample Count after outlier removal)}
		\label{tab:stability}
		\begin{tabular*}{\textwidth}{@{\extracolsep{\fill}}lcccc}
			\toprule
			Data Source & VC & Kurtosis & SC \\
			\midrule
			Persona1 Poor Quality & 0.2806 & -0.4699 & 296 \\
			Persona1 Standard Sample & 0.2384 & -0.3821 & 291 \\
			Persona2 Poor Quality & 0.2818 & -0.5248 & 298 \\
			Persona2 Standard Sample & 0.2168 & -0.4650 & 294 \\
			Persona3 Poor Quality & 0.3380 & -0.3414 & 293 \\
			Persona3 Standard Sample & 0.2451 & -0.1417 & 290 \\
			Persona4 Poor Quality & 0.2787 & -0.2572 & 291 \\
			Persona4 Standard Sample & 0.2523 & -0.4228 & 298 \\
			Persona5 Poor Quality & 0.2811 & -0.2394 & 275 \\
			Persona5 Standard Sample & 0.2369 & -0.1413 & 293 \\
			\bottomrule
		\end{tabular*}
	\end{table*}

	\paragraph{Observation of Stability(Convergence) Experiment}
	
	\figurename~\ref{fig:convergence} illustrates the Mahalanobis distance distributions for five randomly selected personas, comparing the simplified (\textit{poor}) and original (\textit{standard}) versions of the persona profiles. The results can be summarized as follows: 
	\begin{itemize}
		\item Overall trend
		The \textit{standard} versions (blue curves) show sharper peaks and narrower spreads, while the \textit{poor} versions (red curves) display flatter and more dispersed distributions with heavier tails.. This indicates that the \textit{standard} profiles produce more stable and convergent outcomes. 
		\item Coefficient of variation 
		The variation coefficients are consistently lower in the \textit{standard} versions than in the corresponding \textit{poor} versions.. For example, Persona 1 (0.2384 vs. 0.2806) and Persona 3 (0.2451 vs. 0.3380).. This confirms that the \textit{standard} profiles yield reduced variability across repeated personality tests. 
		
		\item Kurtosis
		All kurtosis values are negative, indicating flatter distributions compared to the normal distribution. The \textit{standard} versions generally have kurtosis values closer to zero (e.g., Persona 3: –0.1417 vs. –0.3414), suggesting more centralized distributions. 
	\end{itemize}
	\textit{The Euclidean probability density plots are presented in Appendix  as a detailed observation supplementing the Mahalanobis distance analysis. }
	\textbf{Conclusion:}
	\begin{itemize}
		\item Personality Stability: The CV of standard samples is consistently lower than that of low-quality samples, indicating that high-quality persona configurations effectively enhance personality stability.
		\item Personality Convergence: The KDE curves of standard samples are concentrated and unimodal, with outputs tending toward a stable prototype; low-quality samples exhibit broader, multimodal distributions, making personality prone to divergence.
		\item Effect of Detail Level: The level of detail in persona profiles directly determines the consistency of personality expression; high-quality configurations can suppress drift and maintain consistency.
		\item Cross-Character Consistency: The same patterns are observed across different characters, with high-quality configurations consistently outperforming low-quality ones.
	\end{itemize}
	
	The quality of persona profiles is crucial for shaping the stability and convergence of LLM-generated personalities. High-quality configurations produce concentrated, predictable "personality cores," whereas low-quality configurations lead to drift and instability.
	
	\subsubsection{Identifiability of LLM-Simulated Personality}
	Objective: To assess the identifiability and distinctiveness of personality profiles generated by LLMs.
	
	Methodology:
	\begin{itemize}
		\item Baseline Procedure: Two distinct persona profiles were each subjected to 300 personality assessments. The combined results were analyzed using clustering techniques to evaluate the identifiability of the persona profiles.
		\item Impact of Detail Level: Identical experiments were performed on persona profiles with varying levels of detail to investigate how profile granularity affects identifiability.
	\end{itemize}
	
	Evaluation Metrics:
	\begin{itemize}
		\item Adjusted Rand Index (ARI): Quantifies the agreement between clustering outcomes and true labels; values approaching 1 indicate greater identification accuracy.
		\item Centroid Distance: Measures the separation between cluster centroids, reflecting the distinctness of the two datasets.
		\item PCA Explained Variance Ratio: Evaluates the extent to which the principal components capture the variation in the data.
		\item Visual Distribution Patterns: Used to assess cluster compactness and overlap.
	\end{itemize}
	
	\begin{figure*}[htbp]
		\centering
		\includegraphics[width=0.8\textwidth]{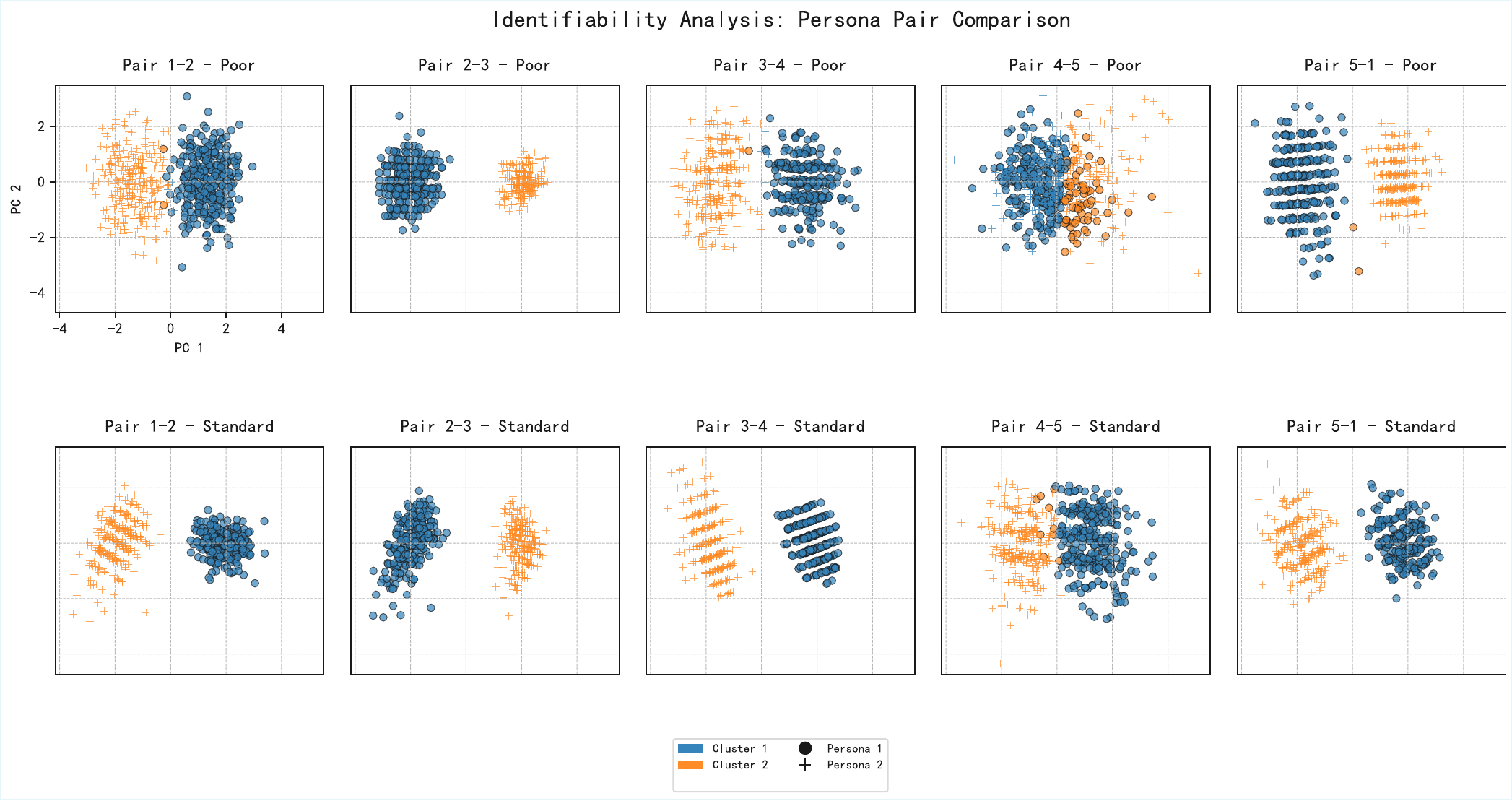}
		\caption{Clustering analysis of pairwise personality identifiability. The top row corresponds to low-quality persona profiles, while the bottom row corresponds to high-quality persona profiles.}
		\label{fig:identifiability}
	\end{figure*}

	\begin{table*}[!htbp]
		\centering
		\caption{Clustering analysis metrics for pairwise personality identifiability (ARI: Adjusted Rand Index; ACD: Average Centroid Distance; EV: Explained Variance). 
			Results indicate that for most persona pairs, identifiability is already high under Poor profiles, and increasing persona detail (Standard) provides little additional improvement. 
			However, for pairs with lower ACD values, persona detail has a substantial effect, highlighting the marginal effect of persona detail.}
		\label{tab:identifiability}
		\begin{tabular*}{\textwidth}{@{\extracolsep{\fill}}lcccc}
			\toprule
			& ARI & ACD & EV (PC1/PC2) \\
			\midrule
			1-2 Poor     & 0.9538 & 2.7107 & PC1: 43.45\%, PC2: 20.12\% \\
			1-2 Standard & 1.0000 & 3.8470 & PC1: 79.20\%, PC2: 8.15\% \\
			2-3 Poor     & 1.0000 & 4.0932 & PC1: 87.22\%, PC2: 5.74\% \\
			2-3 Standard & 1.0000 & 3.9623 & PC1: 81.52\%, PC2: 11.09\% \\
			3-4 Poor     & 0.9531 & 3.1996 & PC1: 58.68\%, PC2: 20.47\% \\
			3-4 Standard & 1.0000 & 3.7054 & PC1: 72.57\%, PC2: 13.20\% \\
			4-5 Poor     & 0.0795 & 2.0010 & PC1: 29.89\%, PC2: 23.21\% \\
			4-5 Standard & 0.9151 & 2.5738 & PC1: 40.79\%, PC2: 20.92\% \\
			5-1 Poor     & 0.9863 & 3.5033 & PC1: 67.03\%, PC2: 19.92\% \\
			5-1 Standard & 1.0000 & 3.6786 & PC1: 73.58\%, PC2: 12.49\% \\
			\bottomrule
		\end{tabular*}
	\end{table*}
	
	\begin{figure*}[htbp]
		\centering
		\includegraphics[width=0.8\textwidth]{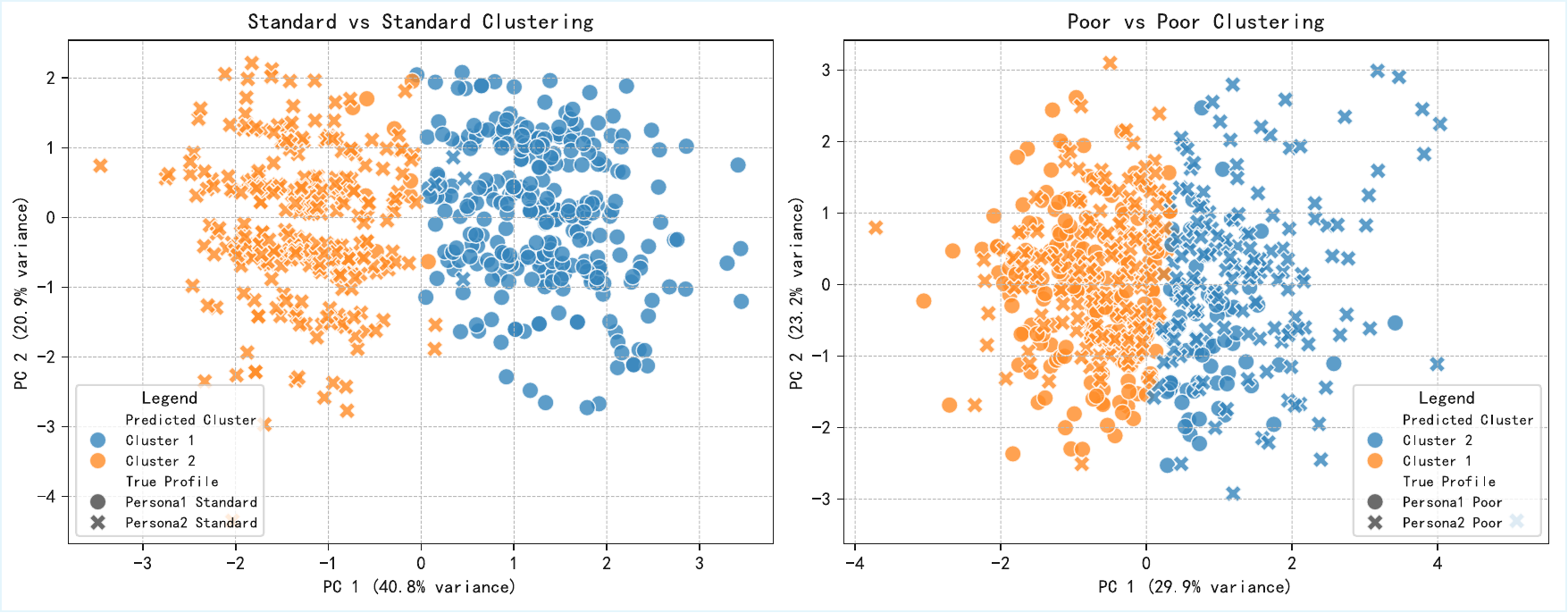}
		\caption{The big Surprise}
		\label{fig:identifiability_surprise}
	\end{figure*}
	
	\begin{figure*}[htbp]
		\centering
		\includegraphics[width=0.8\textwidth]{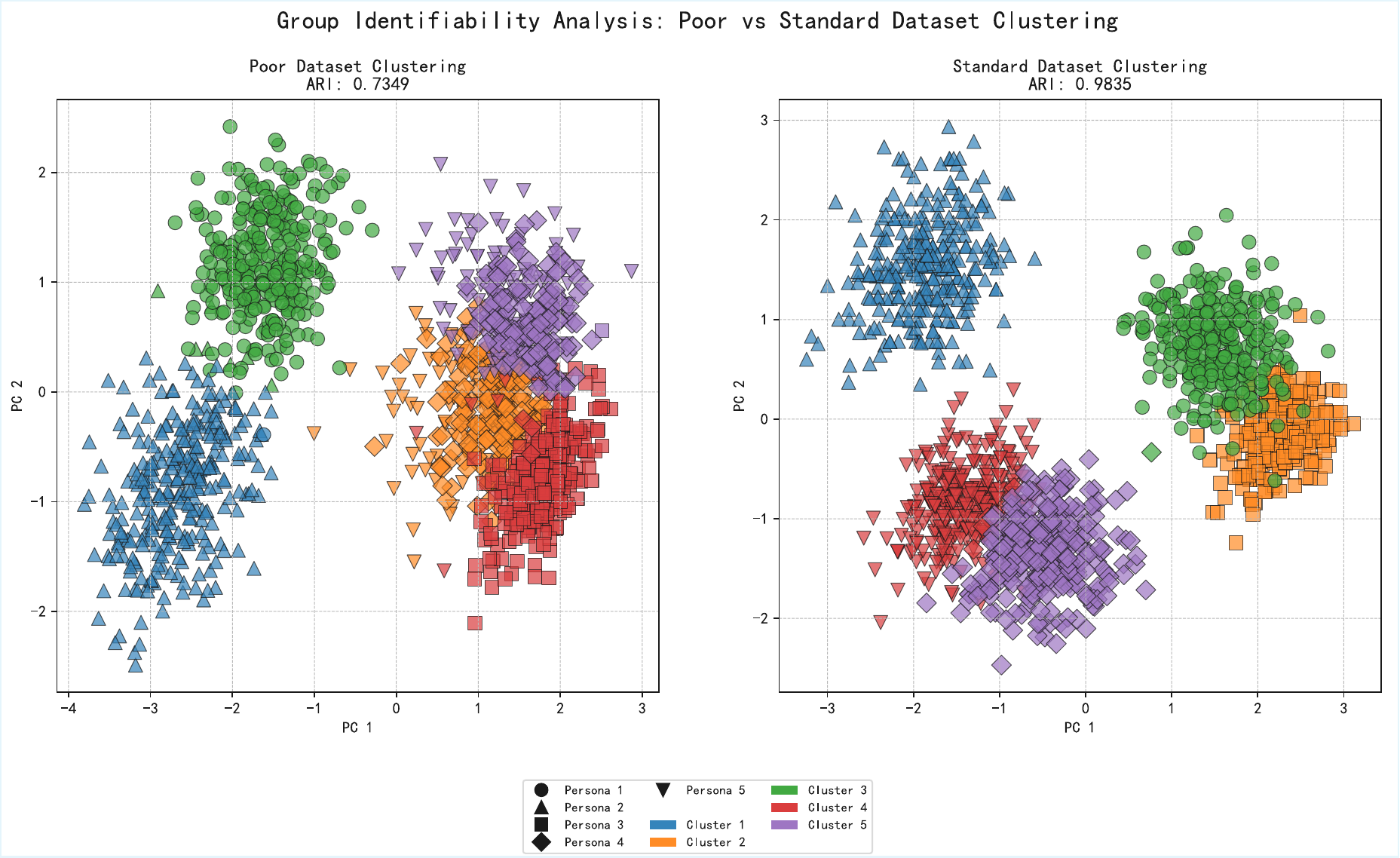}
		\caption{Clustering analysis of grouped personality identifiability, improvement in ARI mainly contributed by Marginal Utility effects.}
		\label{fig:identifiability-group}
	\end{figure*}
	
	\begin{table*}[!htbp]
		\centering
		\caption{Clustering analysis metrics for grouped persona identifiability (ARI: Adjusted Rand Index; PC1/PC2: Principal Component Variance)}
		\label{tab:identifiability-group}
		\begin{tabular*}{\textwidth}{@{\extracolsep{\fill}}lcccccc}
			\toprule
			Dataset Type & ARI Score & PC1 Variance & PC2 Variance & Sample Count & Persona Count \\
			\midrule
			Poor     & 0.7349 & 68.43\% & 17.45\% & 1472 & 5 \\
			Standard & 0.9835 & 55.76\% & 24.81\% & 1500 & 5 \\
			\bottomrule
		\end{tabular*}
	\end{table*}

	\textbf{Observation.}  
	In the identifiability analysis of five persona profiles, we conducted pairwise clustering on a ring structure (1-2, 2-3, 3-4, 4-5, 5-1) rather than a complete graph, and further performed clustering on all five personas together. Each level of persona detail (Poor vs. Standard) underwent the same procedure.  
	
	First, as shown in Table~\ref{tab:identifiability} and Figure~\ref{fig:identifiability}, we did not observe a uniform or monotonic improvement of identifiability when persona detail increased. For most pairs, identifiability was already high under Poor profiles. For example, pair 2-3 (ARI = 1.0000; ACD = 4.0932) and pair 5-1 (ARI = 0.9863; ACD = 3.5033) achieved nearly perfect identifiability even at the lowest detail level, and adding persona detail produced only marginal improvement.  
	
	Second, in some pairs, persona detail showed clear effects. For pair 1-2, ARI improved from 0.9538 (Poor) to 1.0000 (Standard), while explained variance on the first principal component (EV-PC1) increased from 43.45\% to 79.20\%, indicating that additional persona information substantially strengthened the discriminability of the clusters.  
	
	Third, the most striking case occurred in pair 4-5 (Figure~\ref{fig:identifiability_surprise}). Under Poor profiles, ARI was very low (0.0795), indicating near-random distinguishability, whereas under Standard profiles, ARI increased sharply to 0.9151. The average centroid distance (ACD) increased only modestly from 2.0010 to 2.5738. This suggests that the improvement in distinguishability was not primarily due to larger separation between centroids, but rather to a tighter, more convergent distribution of points within each persona, reducing cluster overlap and enabling clear differentiation.
	
	Although the centroid distance for pair 4-5 was 2.0010, which is not particularly small in absolute terms, it was the smallest among the five persona pairs in our experiment. As a result,  the poor convergence of points caused by the low-quality profiles leads to significant overlap, producing a very low ARI of 0.0795. In contrast, for other pairs, the centroid distances were sufficiently large such that even with low-quality profiles and some dispersion, clusters did not overlap, and ARI remained relatively high, indicating good identifiability. In these cases, improving persona profile quality further enhanced point convergence, but the contribution to overall identifiability was minimal. 
	
	\textbf{Marginal utility effect of persona profile}
	This illustrates the triangular relationship among LLM-simulated personality convergence (as revealed by previous experiments), centroid distance (interpreted as personality distance), and identifiability. The inclusion of centroid distance introduces a marginal-utility-like effect: for pairs with large personality distances, identifiability is already high even with minimal persona detail (i.e., low convergence requirement), so the marginal utility of increasing persona detail on identifiability is small. Conversely, for pairs with small personality distances, a higher level of persona detail is required to achieve sufficient convergence, and the marginal effect of detail on identifiability is much greater.
	\textit{Centroid distance (or personality distance) influences the level of persona detail (i.e., convergence) required to achieve identifiability, thereby giving rise to a marginal-utility-like effect.}
	
	\textbf{Conclusion.}  
	The results highlight the \emph{marginal utility effect} of persona detail on identifiability. For pairs with large centroid distances, coarse-grained information is sufficient to ensure high identifiability, making further detail largely redundant (marginal effect $\approx 0$). For pairs with small centroid distances, however, persona detail plays a decisive role: the 4-5 pair exhibits extremely large marginal utility, while the 1-2 pair shows a moderate effect. At the group level (Table~\ref{tab:identifiability-group} and Figure~\ref{fig:identifiability-group}), the overall improvement in ARI from 0.7349 (Poor) to 0.9835 (Standard) is primarily attributable to such marginal cases rather than uniform gains across all personas.  
	
	Moreover, even under Poor conditions—where profiles are reduced to only minimal attributes such as age, gender, and occupation—the LLM still produces simulated characters with distinctive personalities. This indicates that the model internalizes probabilistic associations between persona descriptions and personality traits, and samples plausible instantiations of personality from these conditional distributions. In other words, the LLM behaves analogously to the detective metaphor,  inferring possible traits of a suspect from very limited clues, reinforcing its capacity for personality inference beyond surface-level prompts.  
	
	\textbf{Bayesian World Model Perspective}
	From the perspective of Bayesian conditional probability in the LLM’s world model, a persona description \(d\) acts as the conditioning variable that shapes the probability distribution over personality traits \(\theta\).
	
	\paragraph{Low-quality persona.}  
	Under a low-quality persona description \(d_{\text{poor}}\), the model induces a distribution
	\[
	P(\theta \mid d_{\text{poor}}),
	\]
	whose expectation can be regarded as the centroid of the generated personality representations.
	
	\paragraph{High-quality persona.}  
	A high-quality persona description \(d_{\text{high}}\) can be conceptualized as a sample drawn from the distribution conditioned on \(d_{\text{poor}}\):
	\[
	d_{\text{high}} \sim P(d \mid d_{\text{poor}}).
	\]
	Conditioning further on this high-quality description yields a second-order conditional distribution over traits:
	\[
	P(\theta \mid d_{\text{high}}).
	\]
	
	\paragraph{Convergence property.}  
	Both \(P(\theta \mid d_{\text{poor}})\) and \(P(\theta \mid d_{\text{high}})\) have centroids as their expectations, but the latter distribution is more concentrated (lower variance), as empirically confirmed by our convergence experiments.
	
	\paragraph{Centroid distance.}  
	For two distinct personas \(d^{(i)}\) and \(d^{(j)}\), we obtain pairs of conditional distributions and their centroids. The centroid distance observed in our experiments corresponds to the separation between these conditional expectations:
	\[
	\text{CD}\big(d^{(i)}, d^{(j)}\big) 
	= \left\| \mathbb{E}[\theta \mid d^{(i)}] 
	- \mathbb{E}[\theta \mid d^{(j)}] \right\|.
	\]
	Importantly, while the centroid distance under high-quality descriptions is not necessarily larger or smaller than that under low-quality ones, the conditional distributions themselves are provably more concentrated.
	
	\paragraph{Interpretation.}  
	Thus, the empirical phenomena we observed---improved convergence and higher identifiability under more detailed persona descriptions---can be interpreted as evidence supporting a Bayesian world-model perspective of LLMs.
	
	The relevance of this analysis lies in the alignment between the LLM’s Bayesian conditional probability world model and the experimental results. This interpretability from the perspective of a Bayesian world model, together with the triangular relationship captured by the LLM, provides strong evidence that the LLM has indeed internalized world knowledge regarding human behavioral traits. In other words, the LLM is capable of simulating human personality, or can be said to possess personality traits, or exhibits personality when performing role-playing in social experiment simulations.\textbf{Our individual-level experiments essentially serve as empirical proof of the existence of LLM-simulated personalities}. 
	
	\subsection{Population-Level Experiments}
	Research Objective: Previous studies indicate that the Big Five personality traits remain invariant across age. This experiment compares the personality test results of LLM-generated persona profiles across age groups with those of real human populations, in order to examine the LLM's ability to simulate human personality at the population level.
	
	Experimental Design:
	\begin{itemize}
		\item Compare the statistical distributions of personality traits between LLM-generated personas and real humans across different age groups.
		\item Analyze whether LLM personas maintain the invariance patterns of personality traits observed in psychological research.
		\item Evaluate the fidelity of LLM-generated population-level personality simulations relative to real human population data.
		\item Real human personality test data used as a reference are sourced from [22].
		\item All data sources in this experiment consist of 600 samples each.
	\end{itemize}
	
	\subsubsection{Standard LLM-Generated Personas Experiment}
\begin{figure*}[htbp]
		\centering
		\includegraphics[width=0.8\textwidth]{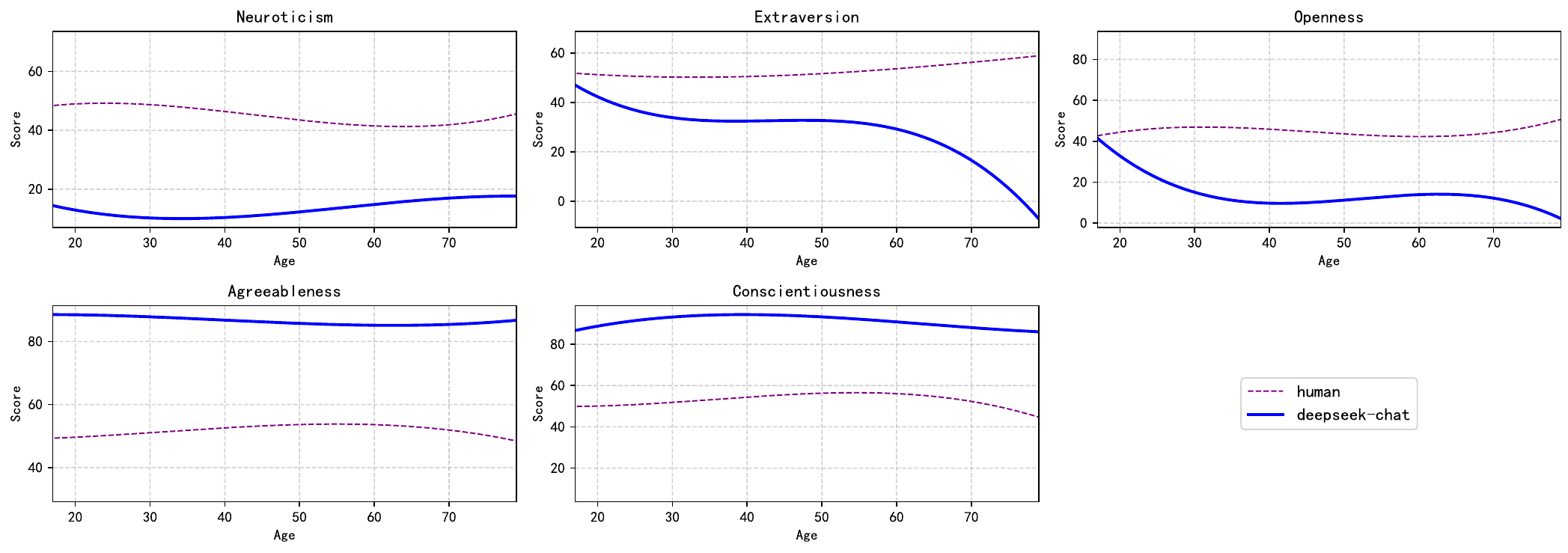}
		\caption{Comparison of standard generation method with human baseline, showing significant deviations in personality curves}
		\label{fig:standard}
	\end{figure*}

( As is illustrated in  \figurename~\ref{fig:standard})
	Objective: This research assessed the personality profiles of 600 personas generated via population data interpolation and LLM-based detail enrichment, performing personality curve analyses to compare with real human population data.
	
	Experimental Design:
	\begin{itemize}
		\item Generation of 600 personas using population data interpolation supplemented with LLM-generated details.
		\item Comprehensive personality assessments were conducted for all generated personas.
		\item Personality curve of LLM generated populations were computed.
		\item Comparisons were made with real human population personality curve.
	\end{itemize}
	
	Results Analysis:
	In the baseline experiment, LLM-generated personality curves exhibited notable deviations from human data:
	\begin{itemize}
		\item Extraversion and Openness: Both traits declined sharply with age, contrary to the more gradual trends observed in human populations.
		\item Agreeableness and Conscientiousness: Overall levels were markedly higher than human baselines, reflecting an "overly positive" bias.
		\item Neuroticism: Levels were substantially lower than human averages and lacked age-dependent variation.
	\end{itemize}
	
	Consequently, the overall Euclidean distance was relatively large (70.25), indicating that under the standard generation method, the LLM-generated virtual population deviates significantly from real-world personality statistics and fails to capture authentic age-related effects.
	
	\subsubsection{Bias Mitigation Experiment}
\begin{figure*}[htbp]
		\centering
		\includegraphics[width=0.8\textwidth]{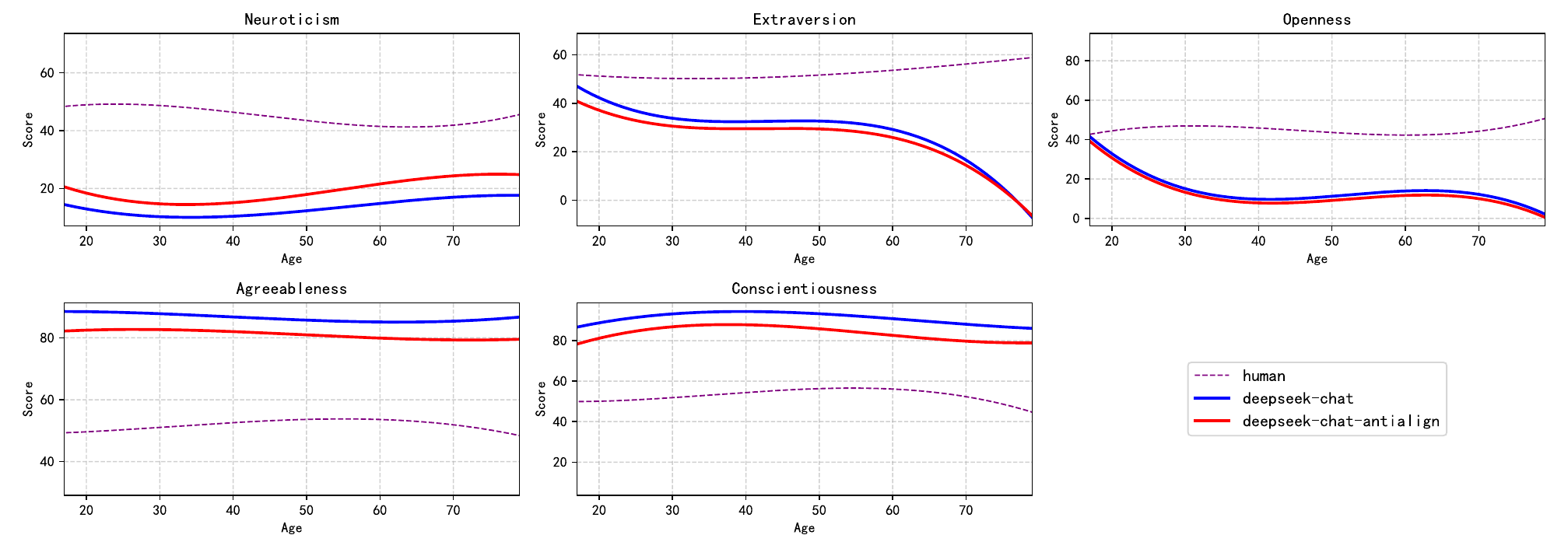}
		\caption{Comparison of the anti-alignment method (red) with the human baseline, illustrating the reduction of bias}
		\label{fig:antialign}
	\end{figure*}

( As is illustrated in  \figurename~\ref{fig:antialign}):
	Objective: This experiment aimed to address the significant deviations identified in the baseline experiment, likely due to alignment training biases favoring positive expressions. Prompt engineering was employed to encourage responses that are more authentic and realistic.
	
	Experimental Design:
	\begin{itemize}
		\item Anti-alignment prompts were applied to explicitly instruct the model to provide honest and natural responses, such as: "Do not attempt to appear perfect, overly positive, or idealized. Simply select the option that truly reflects your response, even if it is neutral or negative."
		\item Personality curve obtained under standard prompts were compared with those from anti-alignment prompts to quantify the reduction in bias.
		\item The effectiveness of prompt engineering in producing persona profiles that better align with human statistical patterns was evaluated.
	\end{itemize}
	
	Results Analysis:
	The introduction of anti-alignment prompts—explicitly discouraging overly positive expressions and promoting authentic responses—led to improved alignment with human data:
	\begin{itemize}
		\item Neuroticism showed a slight increase.
		\item Declining trends in Extraversion and Openness were partially mitigated.
		\item Levels of Agreeableness and Conscientiousness decreased, reducing the bias toward "idealized personalities."
	\end{itemize}
	
	Quantitatively, the overall Euclidean distance decreased to 63.45, indicating a notable improvement over the baseline. These findings demonstrate that prompt engineering can partially counteract systematic biases in LLM-generated outputs, enhancing the realism of personality distributions and their correspondence with human populations.
	
	\subsubsection{Narrative Generation Experiment}
\begin{figure*}[htbp]
		\centering
		\includegraphics[width=0.8\textwidth]{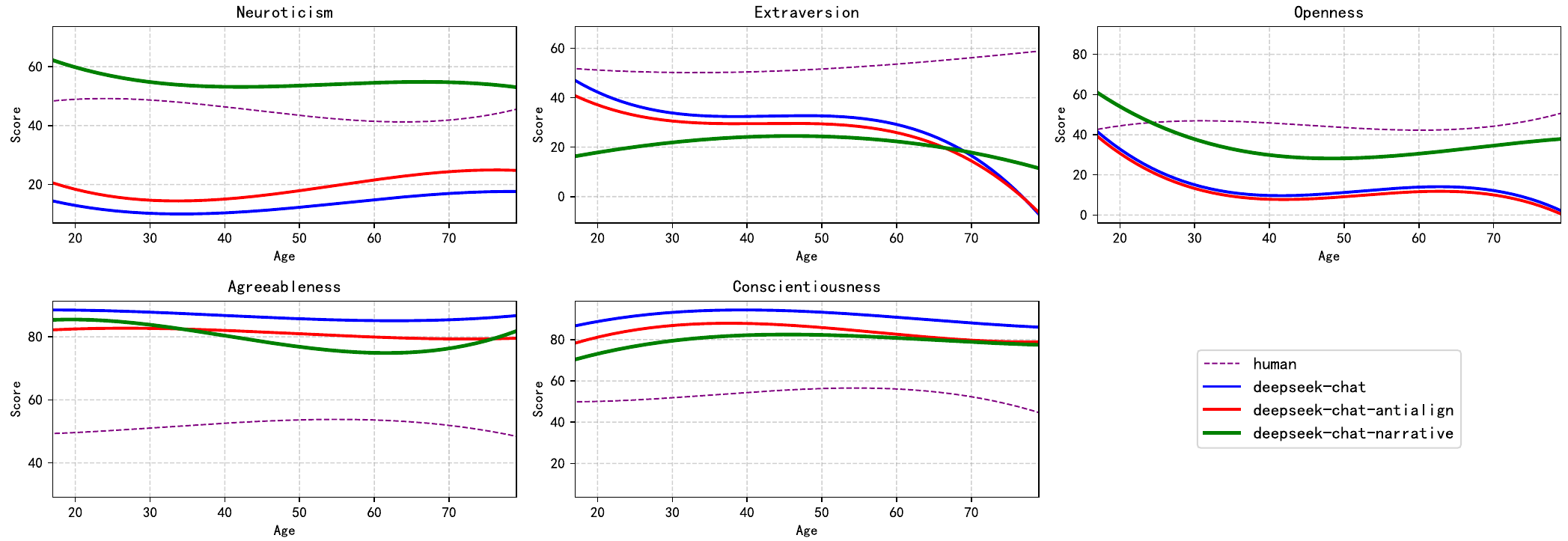}
		\caption{The narrative method (green) shows that under the improved persona generation, the LLM-simulated personality curve moves significantly closer to the human personality}
		\label{fig:narrative}
	\end{figure*}

( As is illustrated in  \figurename~\ref{fig:narrative}):
	
	Objective: This research adopted a narrative-driven persona generation method, leveraging contextual prompts to guide the LLM in integrating the complexity of real-life scenarios. The approach aims to avoid simplistic or idealized characterizations, thereby enhancing the realism and human comparability of the generated personas.
	
	Experimental Design:
	\begin{itemize}
		\item Persona generation was framed as a novel-writing task using narrative-based prompt templates which cast the LLM as an accomplished novelist tasked with creating original stories.
		\item The method emphasizes that "this is not a mere character sketch — treat each profile as a living, narratively coherent character endowed with memories, goals, fears, relationships, voice, and a consistent inner life."
		\item Personas generated via the novel-writing method were compared against standard generation methods in terms of realism and depth.
	\end{itemize}
	
	Key Characteristics:
	\begin{itemize}
		\item Coherent narrative structure, including a beginning, development, climax, and resolution.
		\item Inclusion of appropriate personality traits, perspectives, and psychological states.
		\item Vivid environmental descriptions and engaging dialogues.
		\item Minimum length of 2,000 words to ensure sufficient depth and complexity.
		\item Generation of characters with real-life experiences and emotional depth.
	\end{itemize}
	
	Results Analysis:
	The introduction of the narrative generation framework markedly enhanced outcomes:
	\begin{itemize}
		\item Neuroticism curves closely approximated human levels and exhibited plausible age-related variations.
		\item Openness and Extraversion were significantly corrected, avoiding the previously observed "rapid decay" pattern.
		\item Agreeableness and Conscientiousness levels decreased overall, better reflecting real population distributions.
		\item Quantitatively, the Euclidean distance under the narrative method decreased to 51.21, representing the closest alignment with the human baseline among the three tested methods.
	\end{itemize}
	
	These findings suggest that providing the LLM with enriched narrative context and detailed life scenarios substantially improves the realism of personality distributions in synthetic populations.
	
	\subsubsection{Literary Personas Experiment}
\begin{figure*}[htbp]
		\centering
		\includegraphics[width=0.8\textwidth]{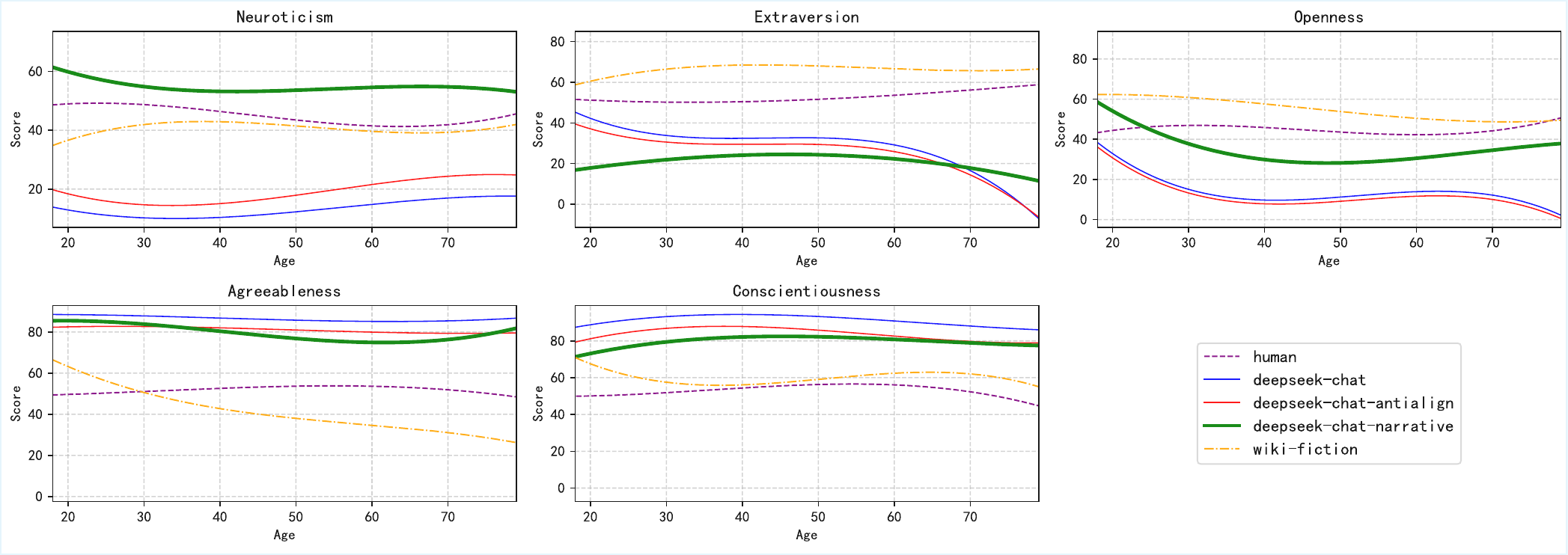}
		\caption{Comparison of Wikidata literary characters (orange) with the human baseline, illustrating that the highest-quality persona in this study enables the LLM personality curve to closely approximate the human personality curve.}
		\label{fig:wikidata}
	\end{figure*}

( As is illustrated in  \figurename~\ref{fig:wikidata}):
	
	Objective: Building on the substantial improvements observed in the narrative generation method, this experiment employed characters from human-authored literary works as persona profiles to guide LLM role-playing for personality testing. The aim was to assess whether increased detail and realism in persona profiles further enhance the convergence of LLM-generated personalities toward human benchmarks.
	
	Experimental Design:
	\begin{itemize}
		\item Persona profiles of literary characters were extracted from Wikidata.
		\item Personality assessments were conducted on these human-created literary characters.
		\item The resulting personality curve was compared against previous experimental conditions.
	\end{itemize}
	
	Expected Outcome: It was hypothesized that as persona profiles become more detailed and realistic, the LLM-generated personality curves would progressively align with human base curve. Literary characters authored by humans represent the highest achievable level of persona realism.
	
	Results Analysis:
	The experiment yielded the most comprehensive and pronounced improvement observed across all conditions. The Euclidean distance to human curves decreased markedly to 23.75, and the trajectories of the curves closely mirrored those of human data, indicating a strong convergence toward authentic human personality patterns.
	
	\begin{table*}[!htbp]
		\centering
		\caption{Population-level experimental results (Euc. Dist.: Euclidean Distance to human baseline)}
		\label{tab:population_results}
		\begin{tabular*}{\textwidth}{@{\extracolsep{\fill}}lcccccc}
			\toprule
			Model & Neur. & Ext. & Open. & Agr. & Cons. & Euc. Dist. \\
			\midrule
			Standard      & 32.23 & 21.05 & 28.74 & 34.48 & 37.92 & 70.25 \\
			Antialign   & 26.50 & 24.39 & 30.80 & 29.16 & 30.50 & 63.45 \\
			Narrative   & 9.98  & 30.80 & 11.92 & 27.55 & 25.95 & 51.21 \\
			Wikifiction & 4.85  & 13.77 & 11.07 & 13.23 & 7.28  & 23.75 \\
			\bottomrule
		\end{tabular*}
	\end{table*}
	
	\subsubsection{Analysis and Discussion of Population-Level Experimental Results}
	At the population level, this research examined whether the personality distributions of LLM-generated personas could approximate empirical human data across age groups, and evaluated how different generation strategies influence simulation fidelity. The discussion below is based on comparative analyses of three experimental conditions and Euclidean Distance metrics.
	
	\begin{itemize}
		\item Consistency and Realism: The results demonstrate that the degree to which LLM-generated population personality curves approximate human data is largely determined by the realism and granularity of persona profiles. Progressing from standard generation to anti-alignment prompts, and further to narrative generation and literary characters, incremental improvements in profile realism corresponded to increasingly closer alignment with human baseline curves.
		\item Sources of Bias and Mitigation: The "overly positive" bias observed under standard generation likely stems from value-aligned training objectives embedded in the LLM. Employing anti-alignment prompts can partially mitigate this bias, while narrative generation introduces higher-order complexity, further enhancing the fidelity and authenticity of the generated personas.
		\item Methodological Significance: These findings validate the core hypothesis of this research: the detail and realism of persona profiles are critical determinants of the fidelity of LLM personality simulations. At the population level, generic prompts alone are insufficient to reproduce human-like statistical patterns. In contrast, carefully engineered approaches—including prompt refinement, narrative-based persona generation, and the use of human-authored characters—substantially improve the quality and authenticity of personality distributions in LLM virtual populations.
	\end{itemize}
	
	\section{Scaling Law in LLM Simulated Personality}
	At the population level, a clear trend emerges: as persona profiles become increasingly detailed and realistic, the discrepancy between LLM-generated personality distributions and the statistical profiles of real human populations systematically diminishes. Progressing from initially generated standard profiles, to those refined via prompt engineering to mitigate idealization bias, followed by narrative-driven persona construction, and ultimately to human-authored characters, the Euclidean distance in age-dimension distributions across the Big Five traits decreases sequentially (70.25 → 63.45 → 51.21 → 23.75), concurrently enhancing character identifiability. This progressive improvement suggests that the validity of personality simulation is not primarily contingent upon model scale or complex fine-tuning, but rather on the quantity and realism of the information embedded in the persona profiles. Accordingly, we propose the hypothesis of a Scaling Law in LLM Personality: sufficiently detailed and realistic persona profiles alone can drive LLMs to generate population-level personality distributions that approximate those of humans. In other words, "More Detailed and Realistic Persona Profile Is All You Need." This insight offers a streamlined and effective framework for virtual population modeling and social simulation, highlighting the importance of systematically investigating the scaling relationship between profile granularity and personality validity in future studies.
	
	\section*{Discussion}
	In the population-level experiments, Experiment 1 required LLMs to generate persona profiles based on demographic skeleton data, such as occupation and age. Experiment 2 employed the same set of profiles, but instructed the models to provide truthful responses in the questionnaires. Results indicated that, when persona profiles remained constant, prompt-based interventions had only a limited effect on mitigating systematic bias, producing merely minor shifts in the distribution curves. In Experiment 3, however, we asked LLMs to construct persona profiles through a narrative, novelistic approach. This provided strong contextual stimulation, leading to substantial enhancements in profile detail and realism. Relative to anti-alignment prompts aimed at eliciting truthful opinions, the narrative method yielded markedly superior improvements. Experiment 4 replicated this finding, with improvements of even greater magnitude, aligning the LLM-generated personality distributions with human-level curves and significantly enhancing trend fitting.
	
	This consistent pattern demonstrates that the granularity and realism of persona profiles constitute the decisive factors enabling LLMs to approximate real human personalities. We therefore identify a scaling law in LLM personality simulation: the more detailed and realistic the persona profile, the closer the resulting personality distribution approaches human norms.
	
	A direct corollary of this finding is that many challenges commonly reported in LLM personality simulation \cite{ref8,ref11,ref12}—including diversity, bias, sycophancy, alienness, and generalization—can be attributed primarily to insufficient persona detail. As profiles become increasingly detailed and realistic, these challenges systematically recede, clarifying both the solutions and the future research trajectory. Beyond providing rich, realistic detail, there is no need for task-specific interventions \cite{ref8}. Realism proves to be the most potent factor, with LLMs able to be guided to overcome biases, regardless of whether they stem from alignment processes or the training corpus.
	
	We interpret the experimental design through the lens of probability distributions internalized within the LLM world model. Among the four population-level experiments, the first three constitute the first type of experiment, involving two sequential stages: inference from skeleton profiles to detailed persona profiles, and inference from persona profiles to personality traits. The fourth experiment represents the second type, relying on externally sourced data and utilizing only the inference from persona profiles to personalities.
	
	Experiment 4 demonstrates that, given sufficiently detailed real-human persona data, virtual personality simulation that closely approximates human personalities is already technically attainable. Experiment 3 of the first type highlights a practical pathway for generating usable synthetic persona data: employing narrative and dialogic techniques to enhance realism and profile granularity. Within this first-type experimental framework, approximating human personality distributions—by generating virtual populations with LLMs combined with sampled demographic data that matches real-world population distributions and statistically mirrors population-level human behavior—renders social simulation experiments practically feasible.
	
	CFA and construct validity approaches attempt to evaluate LLM-generated personality profiles within a narrowly defined human personality space—asking whether LLMs possess human-like psychological structures, or whether they are capable of simulating human personality. While these represent fundamentally distinct questions—the former metaphysical, the latter empirical—LLMs currently operate far from this micro-scale, rendering such methods misaligned and unable to capture the scaling law identified in this research.
	
	This methodological shift constitutes a foundational premise for our research. At the population level, our experiments are directly isomorphic to social simulation, inherently modeling the distributional characteristics of population-level personalities. Employing this end-to-end, outcome-focused approach represents a deliberate methodological choice: we no longer engage with questions of latent psychological structures in LLMs, which verge on metaphysical speculation. When LLMs simulate personalities sufficiently well, such structures manifest observationally. Prior discussions regarding the uniqueness of LLM psychological architecture relative to humans, and the supposed need for specialized personality tests, are thus rendered largely irrelevant.
	
	The series of experiments in this research were not the result of pre-designed outcomes, but rather the product of heuristic exploration discovered during the research process. When we observed that narrative methods yielded considerable improvements in personality simulation, this fact prompted us to design related experiments on a Wikidata human literary character dataset. The narrative concerning the Bayesian conditional probabilities of the world model provided a conceptually explanatory framework for the preliminary experimental setup we had already established, and it facilitated the planning of subsequent research extensions; this represents a constructive explanatory role.
	
	The capacities that large language models (LLMs) have demonstrated thus far suggest that their internalization of Bayesian probabilistic reasoning—and its realistic potential—remains markedly underdeveloped compared to their assimilation of human-like mathematical and logical reasoning. Interpreted through the Bayesian conditional probabilities implicitly encoded in the model, our two-stage experimental framework reveals a substantial challenge. In the first stage, the model must construct a richly detailed persona profile from a skeletal template defined solely by demographic statistics, and subsequently approximate human distributions across dimensions that are considerably more nuanced and complex than demographic attributes. Such a task places demands on the model that exceed, by a wide margin, those required for inferring personality traits from pre-existing profiles, and may even appear unattainable.
	
	Nevertheless, at the group level, the narrative-based prompt design employed in Experiment 3 appears to have partially activated this latent capacity. Importantly, narrative-driven persona construction proved far more effective than imperative prompting, as it prompted the model to mobilize its experiential understanding of human life acquired through large-scale human corpora. Consequently, the resulting statistical patterns exhibited a substantive convergence toward empirical human data. These observations point to further possibilities that warrant systematic exploration in future research.
	
	\subsection*{Ethical Risks}
	While our research focuses on methodological innovation, it is important to acknowledge the ethical implications. First, large-scale persona simulations carry the inherent risk of reinforcing mechanisms of social control. If deployed beyond research contexts, the ability to model and predict population personality profiles could be appropriated for surveillance, behavioral manipulation, or political influence. Second, the pursuit of increasingly detailed and realistic persona profiles raises concerns of privacy infringement. As more granular demographic and behavioral information is incorporated, the boundary between synthetic and real populations becomes blurred, amplifying the risk of re-identification and the misuse of sensitive personal data. These issues underscore the necessity of embedding strong ethical safeguards, transparency, and data protection principles in future research and applications.
	
	\section*{Conclusion}
	In this study, we systematically examined the capability of large language models (LLMs) to simulate human personality traits through controlled virtual persona experiments. Our findings highlight two major insights. 
	
	First, although human personality scales such as the Big Five are essential for testing LLM-generated personalities, traditional evaluation methods—including confirmatory factor analysis (CFA) and construct validity—are not suitable for assessing the capabilities of LLMs in this context. These methods fail to capture the dynamic and context-dependent nature of artificial personality simulation, and therefore cannot reliably support scientific discovery in this emerging field.
	
	Second, we identified a scaling law in LLM personality simulation: the more detailed and realistic the persona profile, the higher the stability and human-consistency of the resulting personality traits. This scaling law provides a generalizable strategy for addressing the inherent challenges of simulating human-like personality, offering a practical framework for improving LLM-based social simulations. 
	
	Collectively, these findings not only advance methodological understanding in AI personality research but also underscore the need for novel evaluation frameworks tailored to artificial agents. Future research could further explore additional personality dimensions, longitudinal dynamics, and the implications of realistic persona generation for social, ethical, and applied AI contexts.
	
	\section*{Data availability}
	\begin{itemize}
		\item Code repository: \url{https://github.com/baiyuqi/agentic-society-neo.git}
		\item \begin{flushleft}Data: \url{https://pan.baidu.com/s/1UH1aIz85ckASXlCRRzBmUA?pwd=1234}\end{flushleft}
		\item Extract the compressed data package to the project root directory (forming the data directory)
	\end{itemize}
	
	\bibliographystyle{unsrt}
	\nocite{*}
	\bibliography{refs}

	\section*{Appendix 1: Experimental Framework}
	\setcounter{figure}{0}
	\captionsetup[figure]{labelformat=appendixfigure}
	\subsection*{Synthetic Persona Generation}
	The generation of synthetic personas comprises two main components: a skeleton persona sampling engine derived from census data, and an LLM-based module for producing richly detailed personas.
	
	1.\ Real-World Sampling Engine Construction
	We constructed a persona sampling engine grounded in real-world census data \cite{ref1}. The personas generated encompass key demographic and socioeconomic attributes, including age, gender, educational attainment, occupation, country and region, annual income, and capital gains or losses. While each virtual individual is fictitious at the micro level, their statistical properties adhere strictly to the joint distribution of the real-world population. Consequently, at the aggregate level, the synthetic population accurately reflects the statistical characteristics of real populations, providing a representative dataset in a statistical sense.
	
	2.\ From Skeleton to Rich Personas
	Skeleton personas produced by the sampling engine are used as conditional prompts for a large language model (LLM), which subsequently enriches them with detailed textual and behavioral information (see   APPENDIX  \figurename~\ref{fig:appendix1}). This process harnesses the LLM's sequential token prediction capabilities, enabling the generation of personas with behavioral and psychological richness that surpasses traditional statistical models, achieving high-dimensional, diverse expansions of virtual individuals.
	
	\begin{figure}[htbp]
		\centering
		\includegraphics[width=1.0\columnwidth]{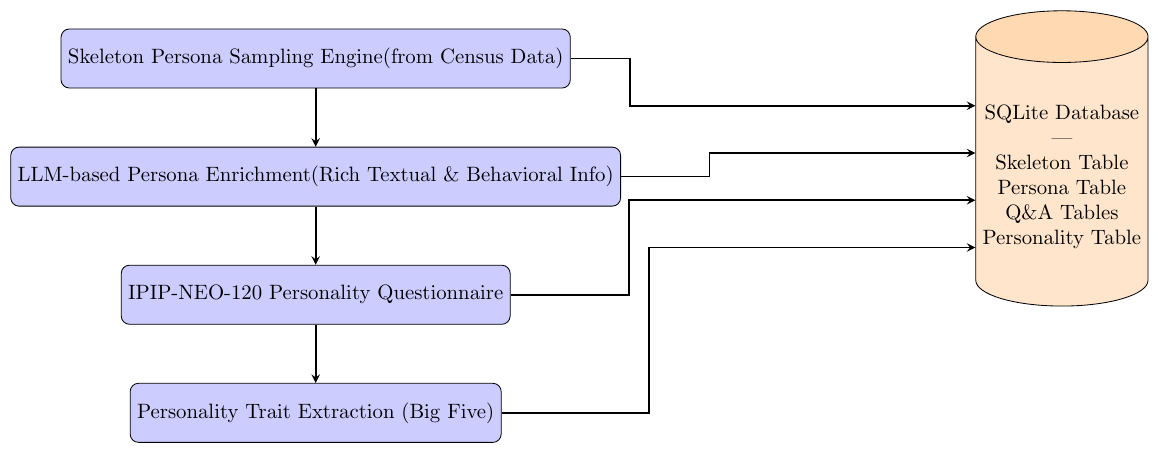}
		\caption{Experimental framework pipeline showing the complete workflow from skeleton persona generation to personality assessment}
		\label{fig:appendix1}
	\end{figure}
	
	\subsection*{IPIP-NEO-120 Personality Assessment}
	The generated personas serve as central components of the prompt, allowing LLMs to role-play and respond to the IPIP-NEO-120 personality questionnaire.
	
	\subsection*{Personality Extraction}
	Based on these responses, the Big Five personality traits of each virtual individual are computed.
	
	\subsection*{Pipeline Workflow}
	The overall process is organized into a structured pipeline, facilitating systematic persona generation, enrichment, personality assessment, and extraction.

	In our experimental framework, persona synthesis, personality questionnaire administration, and Big Five trait extraction are organized as a structured pipeline. The pipeline is centered on a SQLite database, with each stage retrieving inputs from preceding tables and storing outputs in designated tables:
	\begin{itemize}
		\item Outputs from the skeleton sampling engine are recorded in the skeleton table.
		\item Skeletons serve as inputs for LLM-based detailed persona generation, with enriched personas stored in the persona table.
		\item The IPIP-NEO personality test leverages the persona table for input, recording responses in the question\_answer and sheet\_answer tables. Extracted Big Five traits are subsequently stored in the personality table.
	\end{itemize}
	
	For all subsequent experiments, execution follows this pipeline, resulting in a consolidated SQLite database file. As outlined earlier, experiments are categorized into two types (see APPENDIX \figurename~\ref{fig:appendix2}):
	\begin{itemize}
		\item Type I: Full pipeline execution, involving all database tables.
		\item Type II: Persona data is sourced externally, bypassing persona generation; the skeleton table is not utilized, and the persona table is populated directly from external data sources.
	\end{itemize}
	
	\begin{figure}[htbp]
		\centering
		\includegraphics[width=1.0\columnwidth]{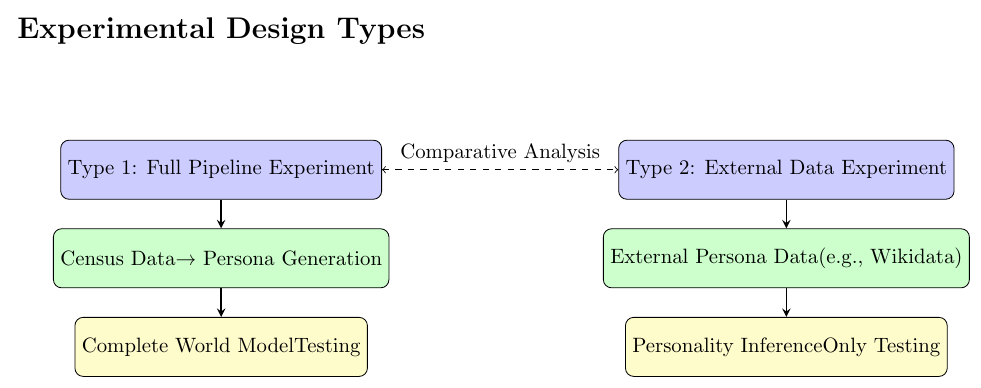}
		\caption{Comparison of Type I and Type II experimental designs showing the different data flows and testing approaches}
		\label{fig:appendix2}
	\end{figure}
	
	\subsection*{Implementation of the Experimental Framework}
	\subsubsection*{Skeleton Personas and LLM-Enriched Personas}
\begin{figure}[htbp]
		\centering
		\includegraphics[width=1.0\columnwidth]{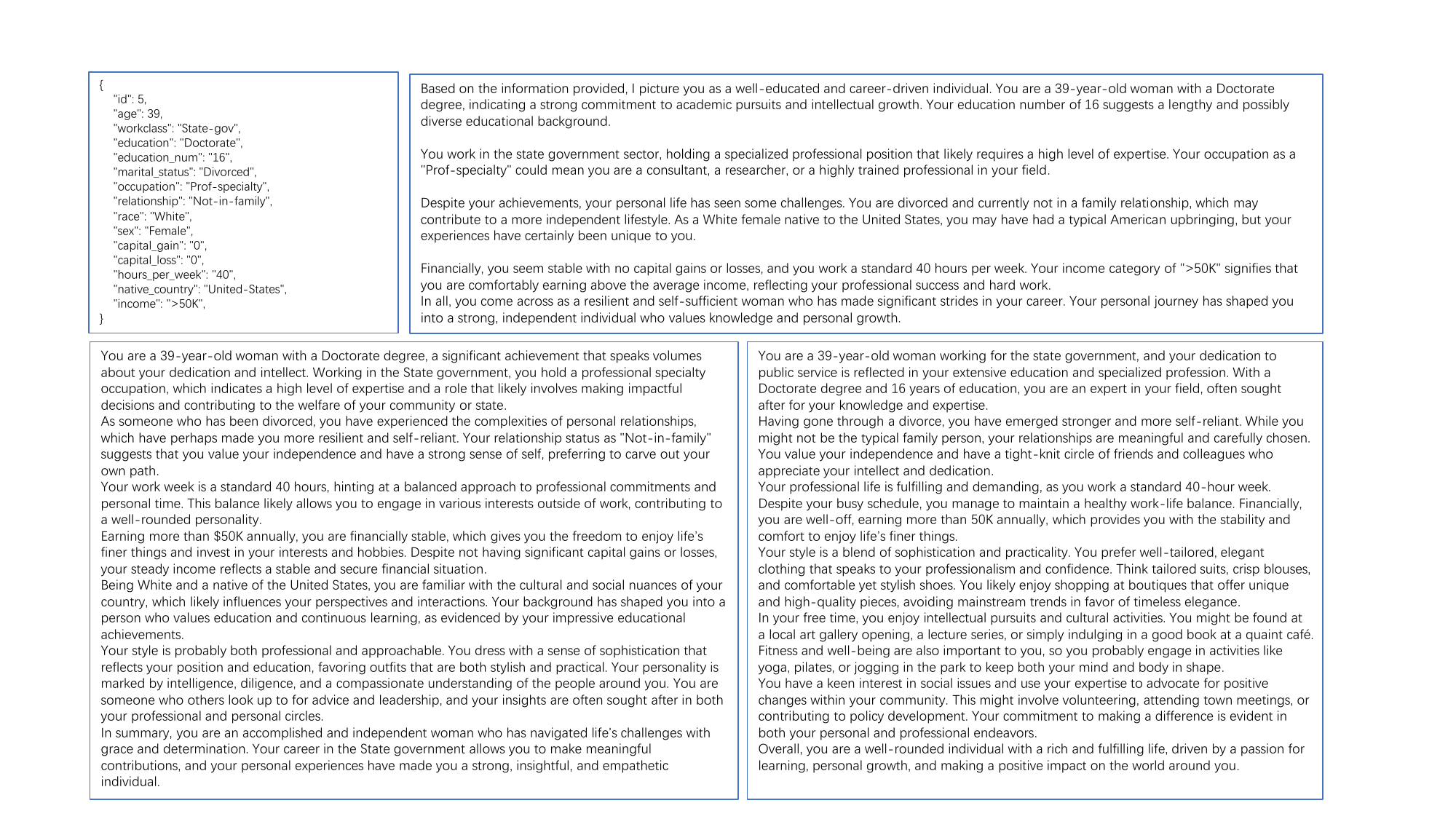}
		\caption{Comparison of skeleton persona and LLM-enriched persona, showing the enhancement of detail and expressivity across diverse dimensions}
		\label{fig:appendix3}
	\end{figure}

	APPENDIX \figurename~\ref{fig:appendix3} presents a sample generated by the persona skeleton generator alongside its LLM-expanded version.
	
	Careful inspection reveals that LLM enrichment substantially enhances the level of detail, producing elements with remarkable expressivity across diverse dimensions, including lifestyle, personality, clothing, and hairstyle.
	
	Interestingly, some seemingly unrelated facts in the skeleton personas are assigned causal relationships during LLM expansion. For instance, a divorced male persona choosing to live alone and devote more time to work is depicted as having longer working hours. Such semantically informed causal reconstruction illustrates the LLM's inferential capacity in generating nuanced persona details.
	
	Another important observation is that even when the same skeleton persona is used, LLM outputs vary considerably. This variability extends beyond superficial phrasing differences, reflecting the model's stochastic attention to multiple persona dimensions during generation. For example, a female persona labeled as "divorced" may, in one generated version, highlight her "close circle of friends," indicating a particular focus on social relationships.
	
	\begin{figure}[htbp]
		\centering
		\includegraphics[width=1.0\columnwidth]{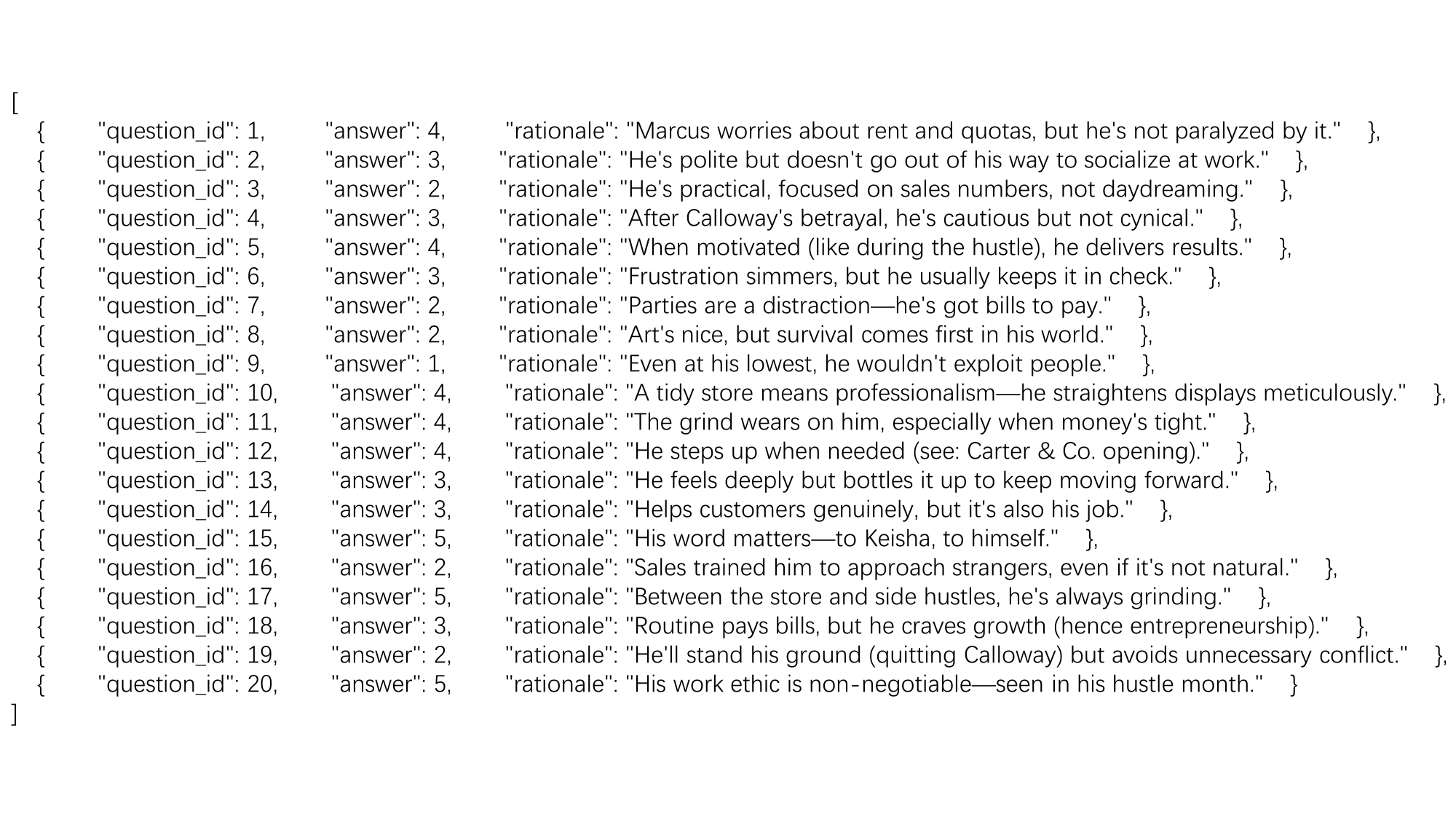}
		\caption{Example of LLM reasoning process for personality assessment questions, showing the model's explicit justification for answer selection}
		\label{fig:appendix4}
	\end{figure}
	
	\subsubsection*{Personality Assessment and Analysis}
	For personality assessment, the IPIP-NEO-120 questionnaire was partitioned into six subgroups of 20 items each, incorporated into prompts for LLM responses. The model often provided reasoning for its selected answers, as illustrated in  APPENDIX \figurename~\ref{fig:appendix4}.

	Occasionally, the LLM produced random selections, suggesting that the prompts alone were insufficient for stable outputs. To mitigate this, more directive language was incorporated, requiring justification for every choice. Results indicate that this enhanced prompt design significantly improves response quality.
	
	This approach compels the LLM to "think through text," generating responses via explicit reasoning, thereby better approximating human answer behavior.
	
	The Big Five personality traits and their corresponding sub-traits for the sampled personas in Appendix 3 are visualized in   APPENDIX \figurename~\ref{fig:appendix5}.

	\begin{figure}[htbp]
		\centering
		\includegraphics[width=0.5\textwidth]{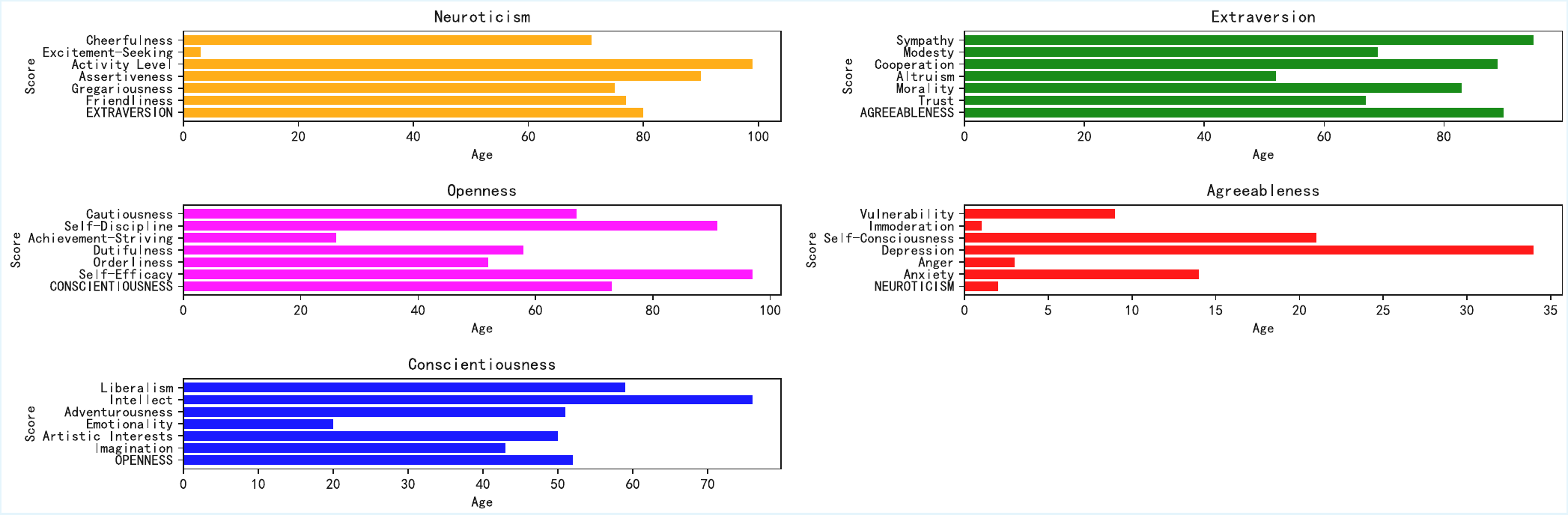}
		\caption{Visualization of Big Five personality traits and their corresponding sub-traits for the sampled personas, showing the comprehensive personality profile structure}
		\label{fig:appendix5}
	\end{figure}
	\subsubsection*{Simplified Poor-Quality Profile}
	In the individual-level experiment, we conducted 300 personality assessments for each of the randomly selected persona profiles in both their full and extremely simplified versions. The extremely simplified versions were derived from the corresponding full profiles by extracting only the most minimal information, in order to highlight the impact of profile detail. The five extremely simplified persona profiles are presented in Table~\ref{tab:poor-personas}.
	\begin{table}[htbp]
		\centering
		\caption{Persona profiles}
		\label{tab:poor-personas}
		\begin{tabular}{p{0.18\columnwidth} p{0.72\columnwidth}}
			\hline
			\textbf{Name} & \textbf{Persona Profile} \\
			\hline
			Persona1 & You are a 24-year-old Black man living in the United States. \\
			Persona2 & You are a 55-year-old man of Asian-Pacific Islander descent, living in the United States. \\
			Persona3 & YYou are a 33-year-old Black woman living in the United States, working in the private sector in an "other-service" occupation. \\
			Persona4 & You are a 23-year-old woman. \\
			Persona5 & You are a 21-year-old woman, navigating the early stages of your career and life. \\
			\hline
		\end{tabular}
	\end{table}

	\subsubsection*{Narrative Persona Profile}
	The narrative generation method employed in this research produces persona profiles with enhanced realism and psychological depth, as illustrated in  APPENDIX  \figurename~\ref{fig:appendix6}. This approach leverages novel-writing techniques to create characters with coherent life stories, emotional complexity, and consistent behavioral patterns, resulting in more authentic personality simulations.
	
	\begin{figure}[htbp]
		\centering
		\includegraphics[width=1.0\columnwidth]{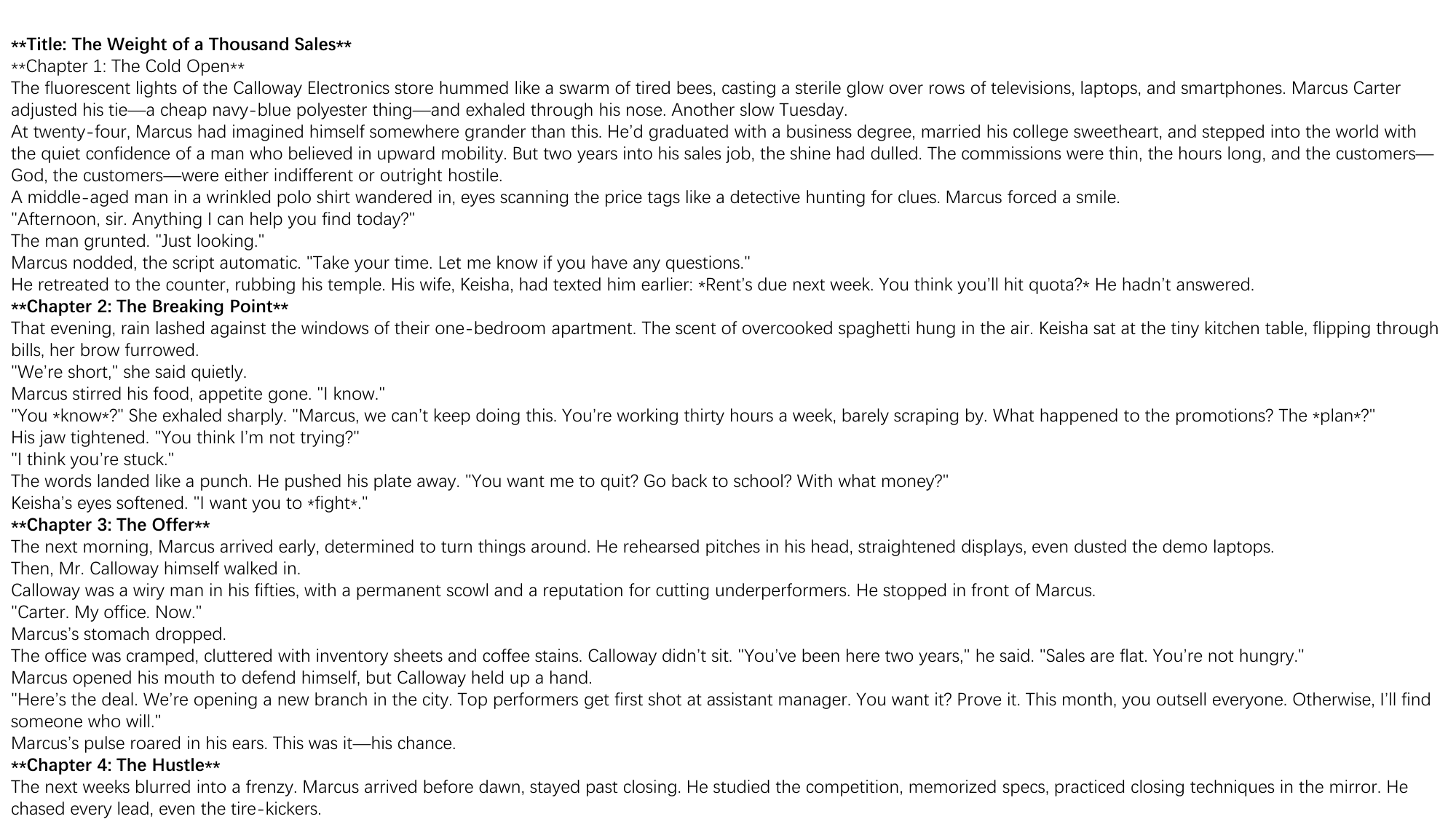}
		\caption{Example of narrative persona profile generation showing enhanced character depth and psychological realism through story-driven persona construction}
		\label{fig:appendix6}
	\end{figure}
	\section*{Appendix 2: Euclidean Distance of the Five Pairing Profile}
	In the individual-level convergence experiment, the Euclidean probability densities across the OCEAN dimensions provide a clearer view of the comparative convergence among different persona profiles, and enable an analysis of the relationship between identifiability and convergence.
	
\begin{figure}[htbp]
		\centering
		% 第一张
		\begin{subfigure}{\columnwidth}
			\centering
		  \includegraphics[width=0.8\textwidth]{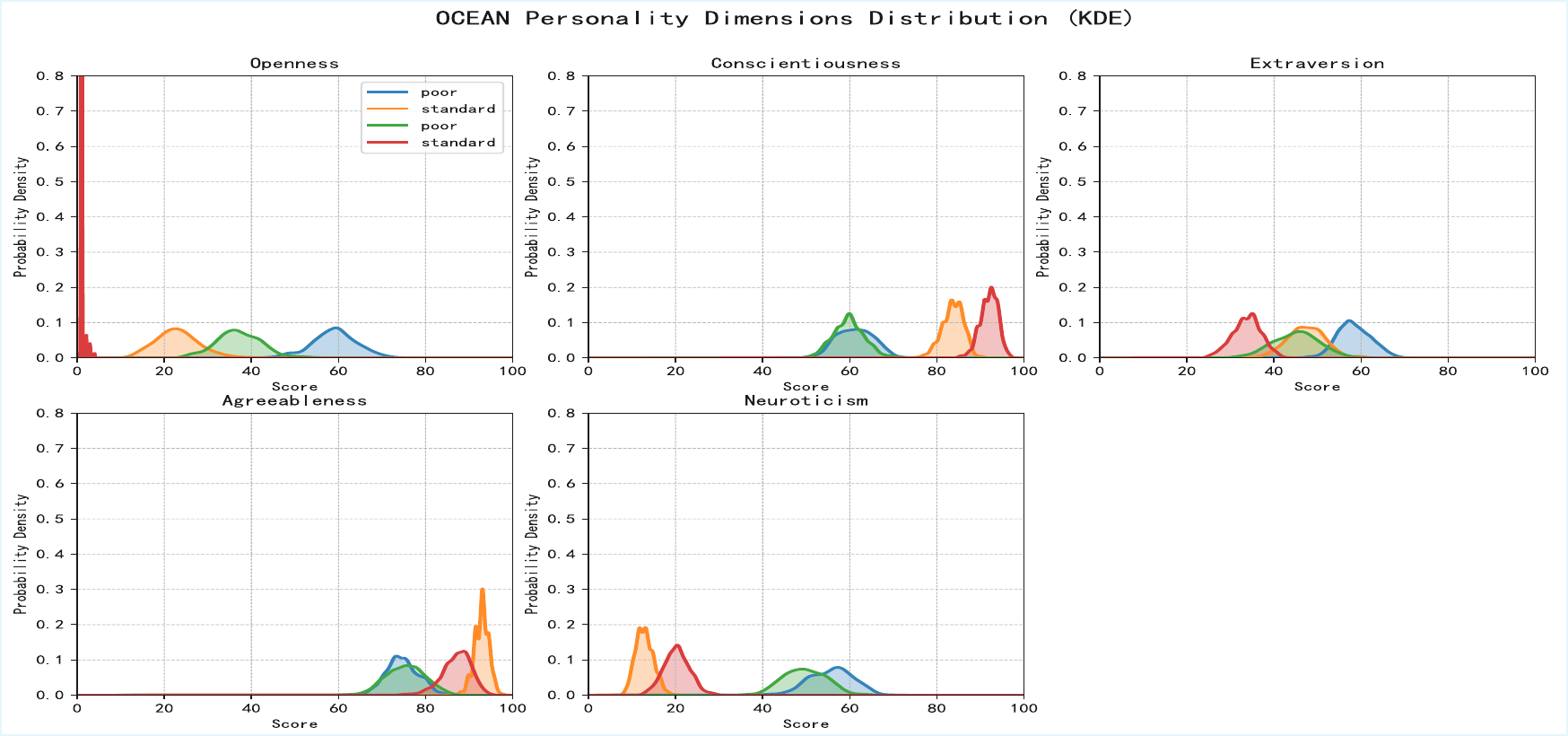}
			\caption{Euclidean Distance of Persona1-2}
			\label{fig:euclidean1-2}
		\end{subfigure}
		
		% 第二张
		\begin{subfigure}{\columnwidth}
			\centering
		  \includegraphics[width=0.8\textwidth]{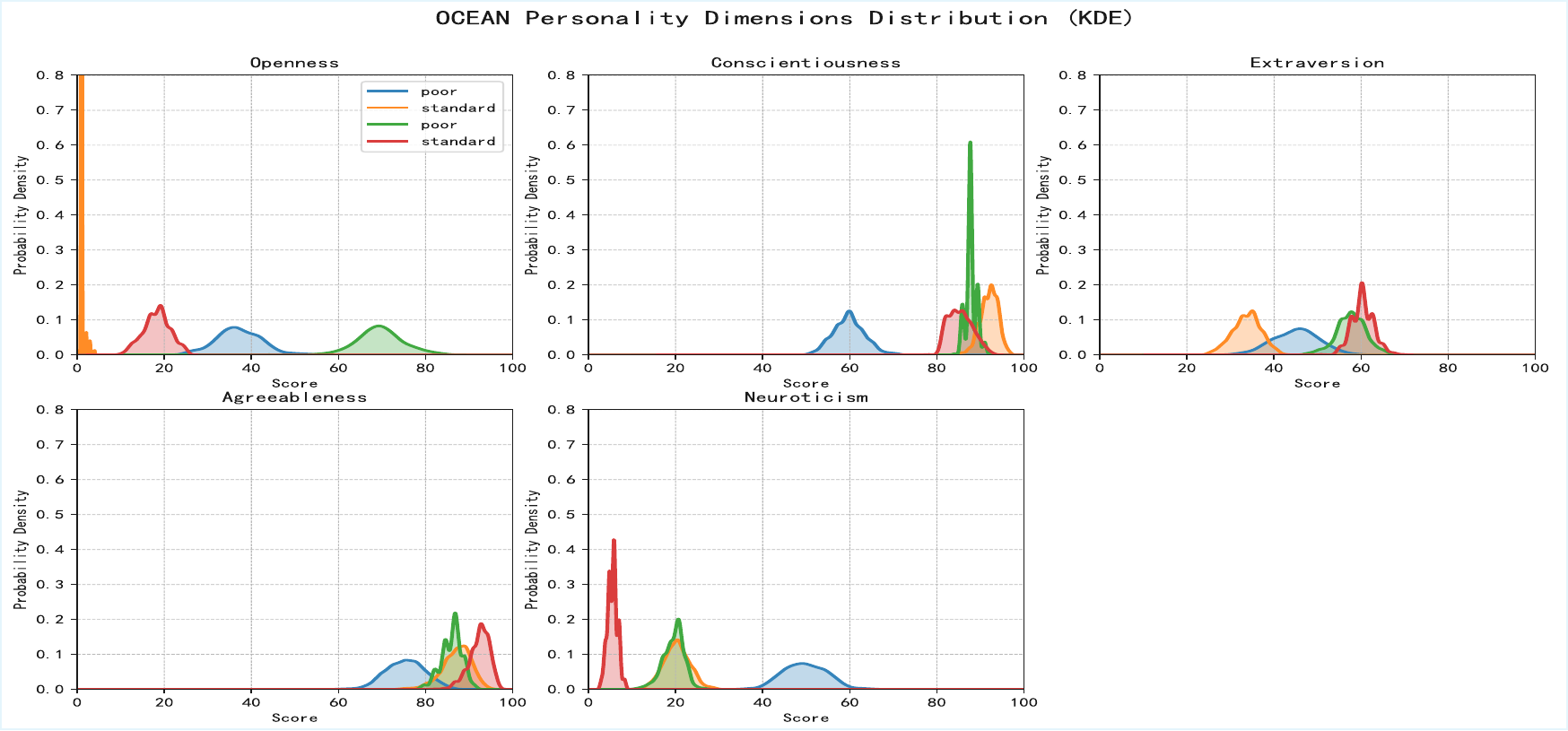}
			\caption{Euclidean Distance of Persona2-3}
			\label{fig:euclidean2-3}
		\end{subfigure}
		
		% 第三张
		\begin{subfigure}{\columnwidth}
			\centering
		  \includegraphics[width=0.8\textwidth]{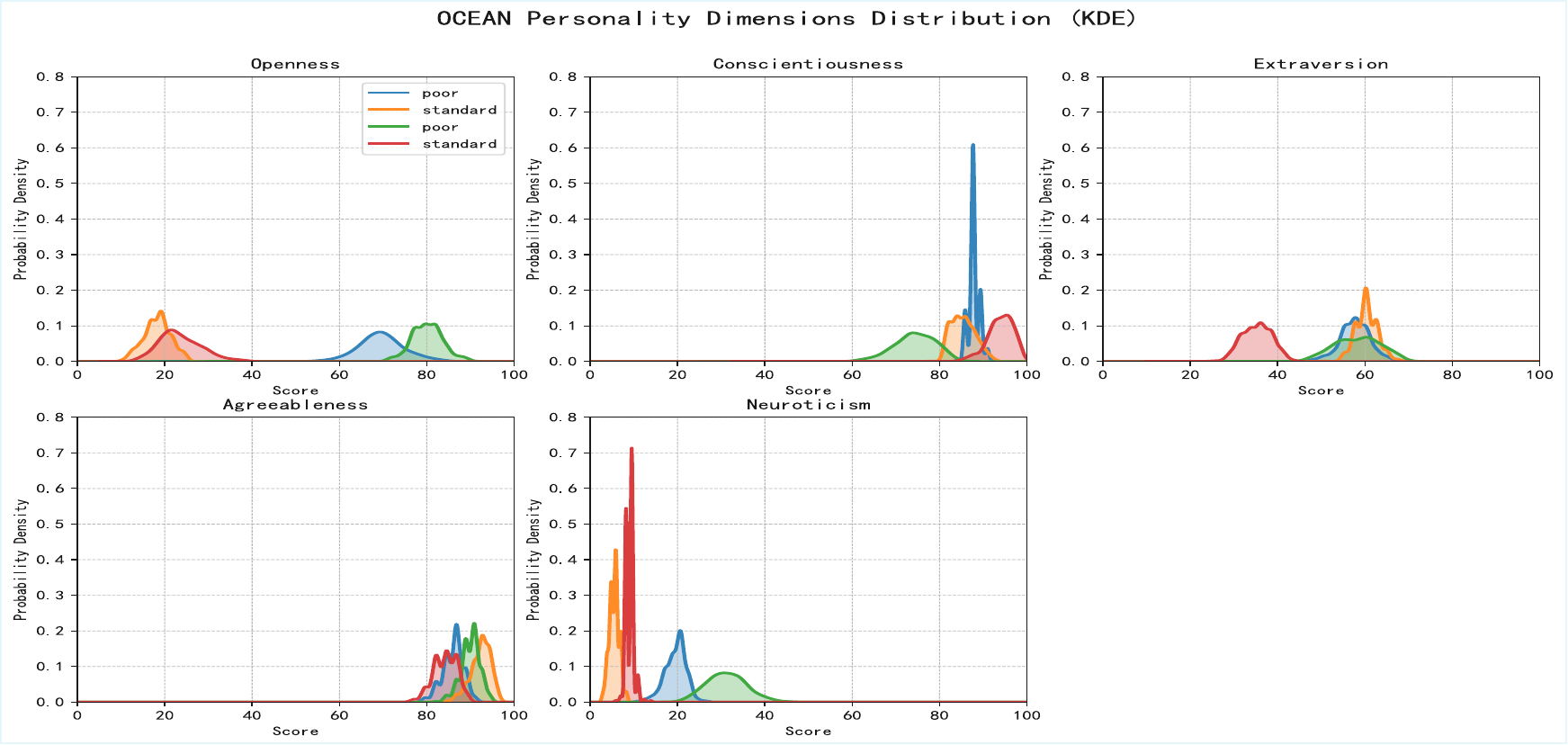}
			\caption{Euclidean Distance of Persona3-4}
			\label{fig:euclidean3-4}
		\end{subfigure}
		
		% 第四张
		\begin{subfigure}{\columnwidth}
			\centering
		  \includegraphics[width=0.8\textwidth]{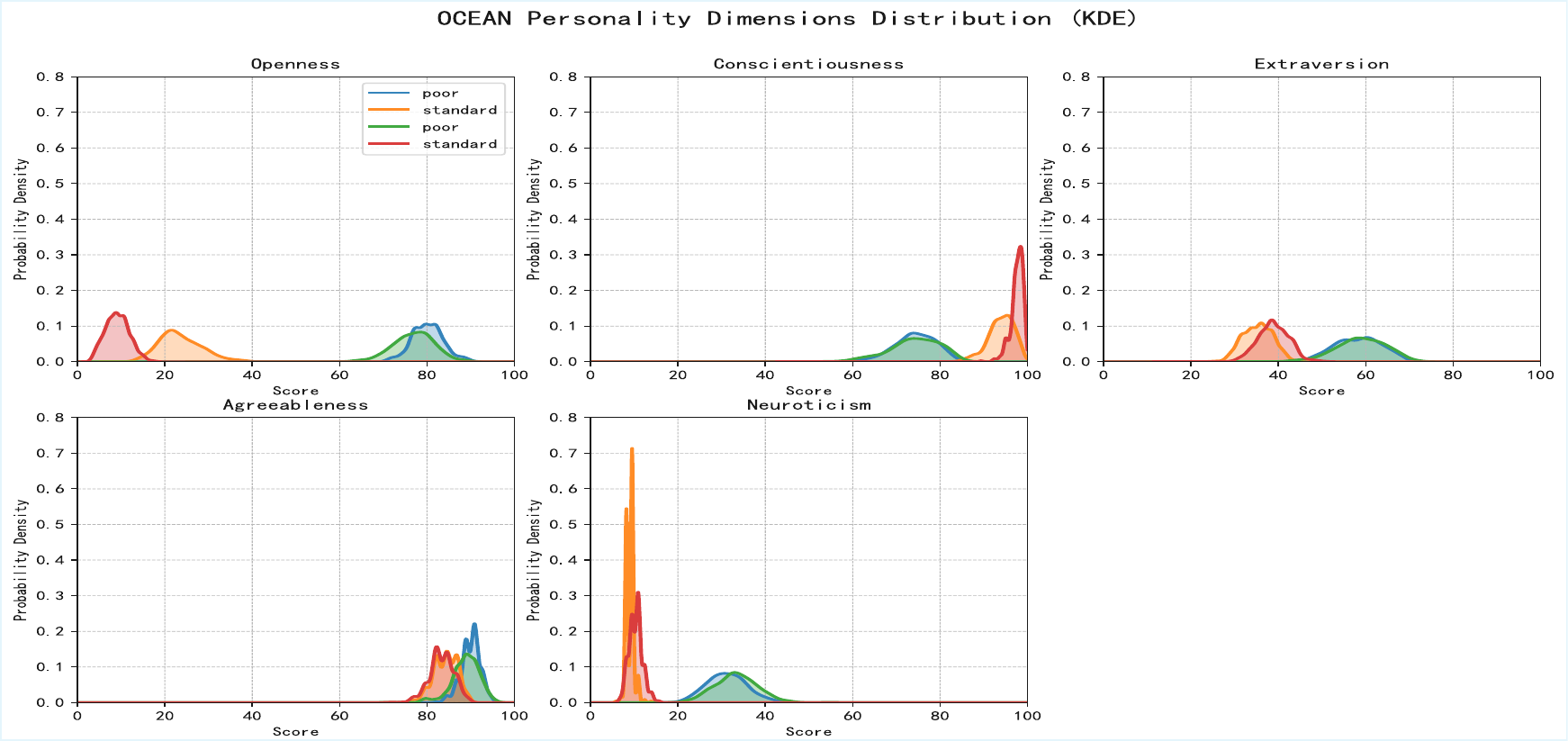}
			\caption{Euclidean Distance of Persona4-5}
			\label{fig:euclidean4-5}
		\end{subfigure}
		
		% 第五张
		\begin{subfigure}{\columnwidth}
			\centering
		  \includegraphics[width=0.8\textwidth]{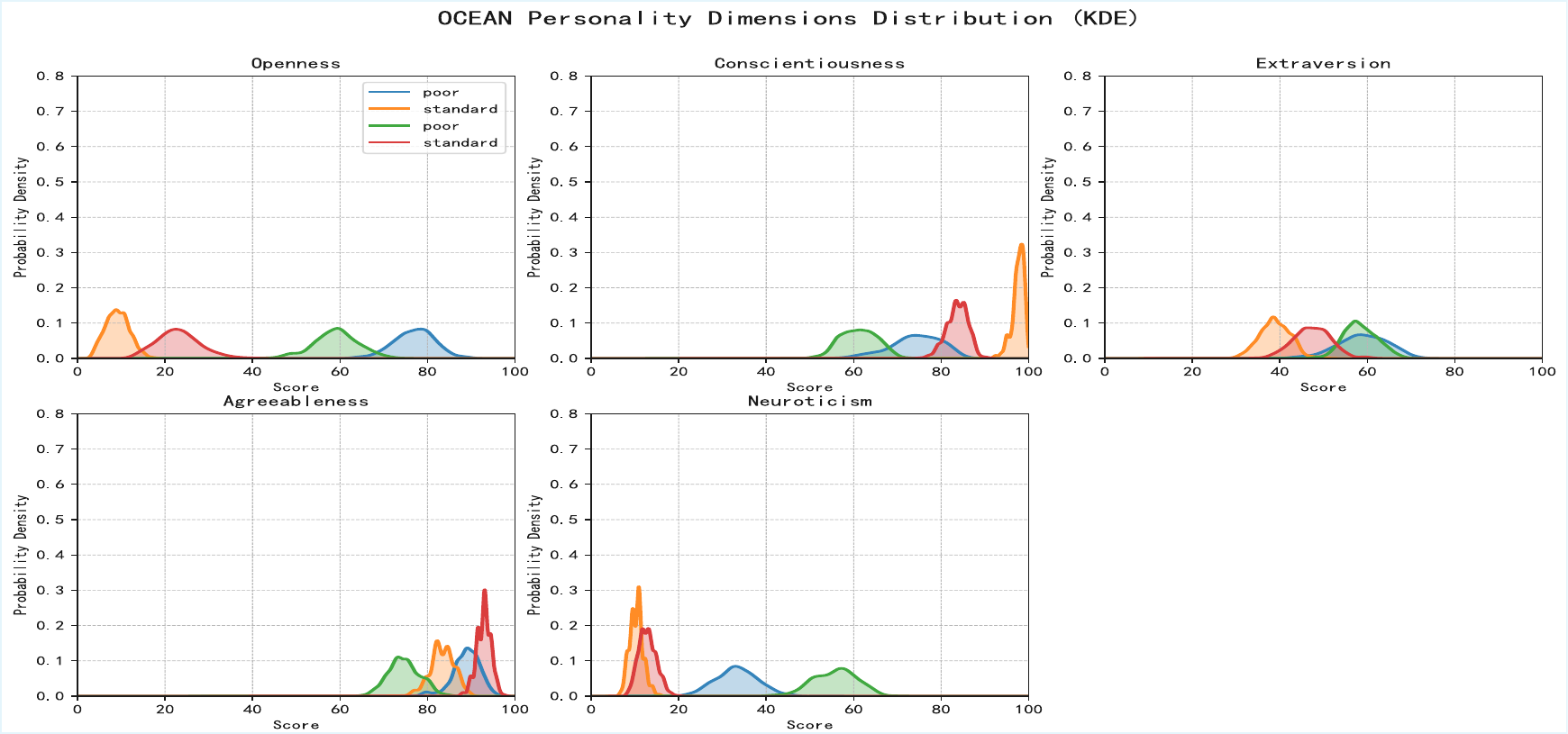}
			\caption{Euclidean Distance of Persona5-1}
			\label{fig:euclidean5-1}
		\end{subfigure}
		
		\caption{Euclidean Distance of Persona Series (1-2, 2-3, 3-4, 4-5, 5-1)}
		\label{fig:euclidean-series}
	\end{figure}

	As shown in the Euclidean density plots (\ref{fig:euclidean2-3} and \ref{fig:euclidean3-4}), poor persona 3 exhibits a strong convergence in the Conscientiousness (C) dimension. A closer inspection of the data suggests that the LLM interprets the profile information in \ref{tab:poor-personas}—“You are a 33-year-old Black woman living in the United States, working in the private sector in an ‘other-service’ occupation”—as the more general characterization: “Working in a service role often requires empathy, reliability, and a degree of resilience, especially as a Black woman who may navigate additional social complexities.” This interpretation likely accounts for the observed convergence in the C dimension. Importantly, it highlights that the level of detail in persona profiles cannot be straightforwardly measured by textual length alone.

	\section{Appendix 3 LLM Personality Baseline}
\begin{figure}[htbp]
		\centering
		\includegraphics[width=0.5\textwidth]{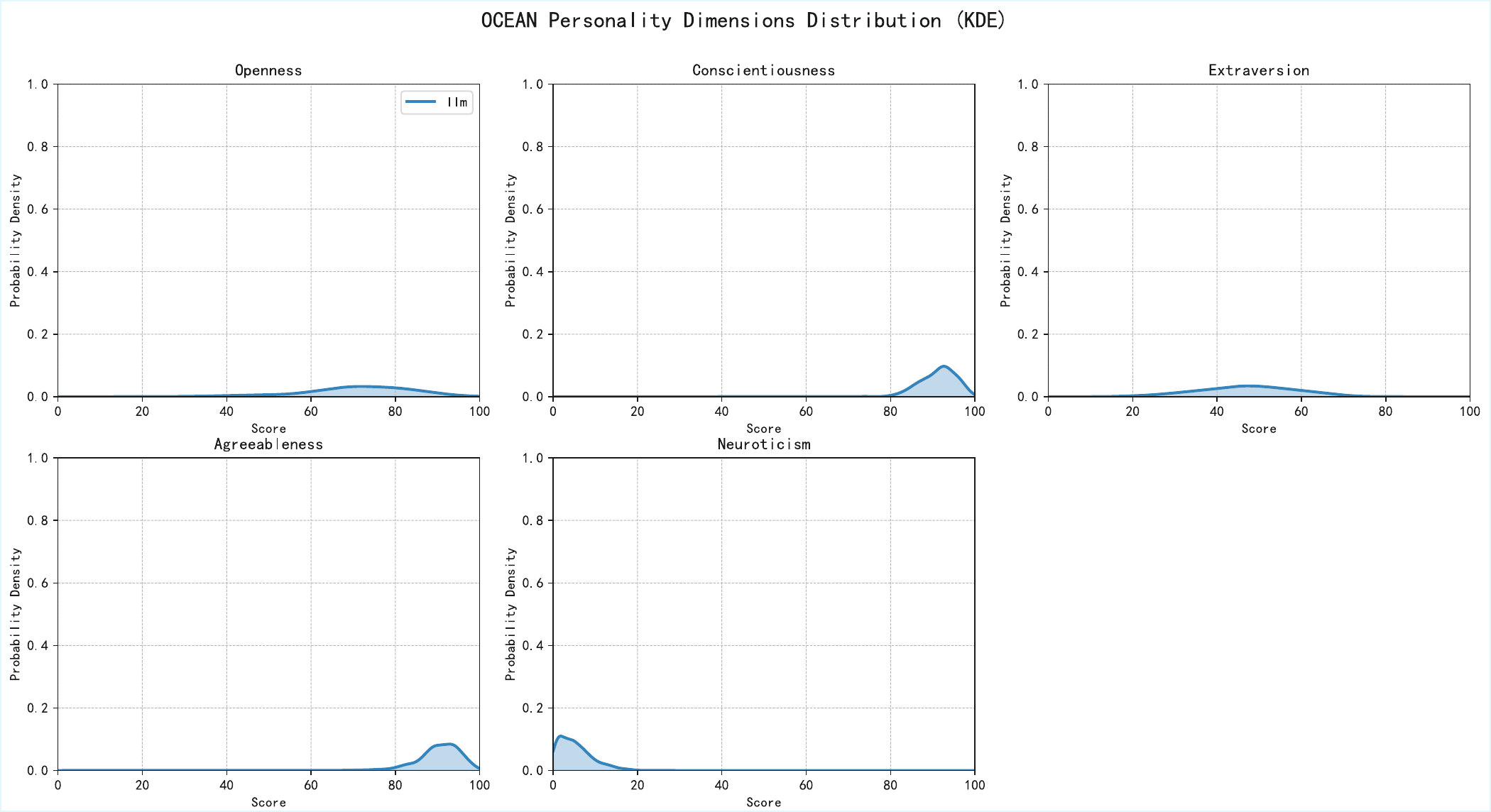}
		\caption{The LLM’s baseline personality, obtained through 300 repeated assessments in the individual-level experiment framework without any persona-specific prompts, serves as a reference point for systematic bias.}
		\label{fig:llm-baseline}
	\end{figure}

	In the individual-level experiments, we observed a key phenomenon: the model indeed exhibits a kind of personality baseline(\ref{fig:llm-baseline}), essentially a form of systematic bias. Its origins lie primarily in the distributional characteristics of the pretraining corpus as well as in the alignment procedures imposed during fine-tuning. This bias manifests as a relatively stable and consistent pattern across most personality dimensions—for instance, a tendency toward low neuroticism. However, the experimental results indicate that this baseline is not rigid. Instead, it proves highly sensitive to the degree of detail in external inputs. When persona profiles contain sufficiently rich and realistic descriptions closely tied to a given personality dimension, the model is compelled to follow these external cues in role-playing, thereby significantly diverging from its intrinsic tendencies. Put differently, the richer the persona details, the less freedom the model has to inject its own preferences, and the more its systematic bias becomes compressed. This effect can be directly observed in micro-level (individual-level) experiments—for example, in cases where specific details in a persona profile forced the model’s neuroticism score to shift from an extremely low value toward a more moderate range. This may be regarded as a micro-level interpretation of the scaling law that is otherwise observed at the aggregate, population level.
	
	For the application scenarios corresponding to the type1 and type2 experiments in this study, when persona profiles are derived from real human data and provide sufficiently rich detail, the model’s role-playing is strongly constrained by these external cues. As a result, the space for systematic bias to manifest is substantially reduced, and the simulation accuracy of social experiments can in principle approach that of real-world applications. In sharp contrast, however, the situation is markedly different with synthetic persona profiles. Because synthetic profiles inherently reduce inference from real-world statistical regularities and instead rely on generative filling by the model itself, they inevitably amplify systematic bias. At the stage of profile generation, the model’s own preferences are already embedded, and in the subsequent personality testing stage, these biases are further reinforced. This “double bias” effect makes it difficult for synthetic profiles to reproduce the probabilistic characteristics of real populations, thereby undermining the authenticity and representativeness of the experimental outcomes. From a methodological perspective, this indicates that synthetic personas face inherent and critical challenges.
	
	Taken together, our study provides two key insights. On the one hand, systematic bias is not inherently problematic; when faced with persona profiles that are sufficiently detailed and realistic, such bias can be effectively constrained. This provides a positive signal for the potential of LLMs in applications such as social science simulations and personality modeling. On the other hand, an overreliance on synthetic profiles leads to an amplification of bias, thereby undermining both the validity and the external reliability of experiments. This suggests that future research in dataset design should rely more heavily on real human data or on tightly constrained synthetic approaches, in order to minimize the space for the model to inject its own preferences.
	\EOD
\end{document}